\documentclass[preprint,notoc]{JHEP3} 
\usepackage{epsfig,multicol}
\usepackage{amssymb}
\usepackage{hep}
\usepackage{graphics}
\usepackage{amsmath,latexsym}

\newcommand{\tb}{\ensuremath{\tan\beta}}
\newcommand{\lag}{\ensuremath{\mathcal{L}}}

\newcommand{\stopp}{\ensuremath{\tilde{t}}}

\newcommand{\mt}{m_t}
\newcommand{\mb}{m_b}

\newcommand{\msusy}{M_{\rm SUSY}}

\newcommand{\bmix}{\ensuremath{\PB_s^0\,-\,\PaB_s^0}}
\newcommand{\ham}{\ensuremath{\mathcal{H}}}

\newcommand{\PDG}{Yao:2006px}

\newcommand{\BurasGabbiani}{Gabbiani:1996hi,Misiak:1997ei}

\newcommand{\Santi}{Bejar:2000ub,Bejar:2001sj,Bejar:2003em}

\newcommand{\FeynArts}{FAFCuser, Hahn:2004fe,Hahn:2005qi}
\newcommand{\bh}{Dimopoulos:2001hw}
\newcommand{\extra}{Hewett:1998sn}
\newcommand{\haber}{Haber:1985rc}

\newcommand{\main}{Guasch:2006hf}
\newcommand{\eilamsis}{Eilam:2006rb}
\newcommand{\GIM}{Glashow:1970gm}

\newcommand{\duncan}{Duncan:1983iq,Duncan:1984vt}

\newcommand{\Liu}{Liu:2004bb}
\newcommand{\cteq}{Pumplin:2005rh}
\newcommand{\lhapdf}{Whalley:2005nh}
\newcommand{\fer}{Ferreira:2005dr,Ferreira:2006xe}

\newcommand{\rau}{Rauchpdh}
\newcommand{\LE}{Diaz-Cruz:1989ub,Eilam:1990zc,Aguilar-Saavedra:2004wm}

\newcommand{\Arhrib}{Arhrib:2004xu,Arhrib:2005nx}
\newcommand{\BGST}{Bejar:2003em}
\newcommand{\Pokorsky}{Gabbiani:1996hi,Misiak:1997ei}
\newcommand{\vegas}{Hahn:2004fe}
\newcommand{\MSSM}{Nilles:1984ex,Haber:1985rc,Lahanas:1987uc}
\newcommand{\mishima}{Endo:2006dm}
\newcommand{\abazov}{Abazov:2006dm}
\newcommand{\bal}{Ball:2003se}

\newcommand{\gab}{Gabbiani:1996hi}

\newcommand{\lat}{Becirevic:2001xt}

\newcommand{\ciuuno}{Ciuchini:2003rj}
\newcommand{\ciudue}{Ciuchini:2006dx}
\newcommand{\process}{\HepProcess{\Pproton \Pproton \HepTo \Ptop \APcharm + \APqt \Pcharm}}
\newcommand{\processgg}{\HepProcess{\Pproton \Pproton(\Pgluon\Pgluon) \HepTo \Ptop \APcharm + \APqt \Pcharm}}
\newcommand{\sq}{\ensuremath{\tilde q}}
\newcommand{\su}{\ensuremath{\tilde u}}
\newcommand{\sd}{\ensuremath{\tilde d}}
\newcommand{\bsg}{\ensuremath{\HepProcess{\Pbottom \HepTo \Pstrange \Pphoton}}}

\title{Single top-quark production by strong and electroweak supersymmetric flavor-changing interactions at the LHC}
\author{
David L\'opez-Val $^{a}$, Jaume Guasch $^{b,c}$ and Joan Sol\`a $^{a,c}$\\
$^{a}\,$ High Energy Physics Group, Dept. Estructura i Constituents de la Mat\`eria,\\
Universitat de Barcelona, Av. Diagonal 647, E-08028
    Barcelona, Catalonia, Spain \\ \\
$^{b}\,$ Gravitation and Cosmology Group,  Dept. F\'isica Fonamental,\\
Universitat de Barcelona,  Av. Diagonal 647, E-08028
    Barcelona, Catalonia, Spain \\ \\
$^{c}$ Institut de Ci\`encies del Cosmos (ICC), UB, Barcelona \\

\email{dlopez@ecm.ub.es}, \email{jaume.guasch@ub.edu},
\email{sola@ifae.es} }

\preprint{UB-ECM-PF 07/27}

\abstract{We report on a complete study of the single top-quark
production by direct supersymmetric flavor-changing neutral-current
(FCNC) processes at the LHC. The total cross section, $\sigma
(\processgg)$, is computed at the $1$-loop order within the
unconstrained Minimal Supersymmetric Standard Model (MSSM). The
present study extends the results of the supersymmetric strong
effects (SUSY-QCD), which were advanced by some of us in a previous
work, and includes the computation of the full supersymmetric
electroweak corrections (SUSY-EW). Our analysis of
$\sigma(\processgg)$ in the MSSM has been performed in
correspondence with the stringent low-energy constraints from
$\bsg$. In the most favorable scenarios, the SUSY-QCD contribution
can give rise to production rates of around $10^5$ events per $100$
$\invfb$ of integrated luminosity. Furthermore, we show that there
exist regions of the MSSM parameter space where the SUSY-EW
correction becomes sizeable. This could be important, especially if
the SUSY-QCD effects would be suppressed. In the SUSY-EW favored
regions, one obtains lower, but still appreciable, event production
rates that can reach the $10^3$ level for the same range of
integrated luminosity. We study also the {possible reduction} in the
maximum event rate obtained from the full MSSM contribution if we
additionally include the constraints from $\PB^0_s-\PaB^0_s$.
However, we treat these restrictions at a different level from the
$\bsg$ ones, due to the higher uncertainties inherent in the
calculation of the matrix element associated to that mixing. In view
of the fact that the FCNC production of heavy quark pairs of
different flavors, such as $\Ptop\APcharm$ or $\APtop\Pcharm$, is
extremely suppressed in the SM, the detection of a significant
number of these events could lead to evidence of new physics -- of
likely supersymmetric origin. }


\begin{document}

\section{Introduction}
\label{sect:introduction}

The forthcoming generation of high energy colliders, headed by the
Large Hadron Collider (LHC) at CERN, and followed by the future
linear collider, depicts an exciting scenario for probing the
existence of physics beyond the Standard Model (SM) of strong and
electroweak interactions\,\cite{Weiglein:2004hn}. Among the possible
discoveries envisioned for the physics at the LHC (some of them of a
rather exotic nature, such as extra dimensions \cite{\extra} and
black-hole production \cite{\bh}), we have the possible confirmation
of the fundamental Higgs mechanism of Electroweak Symmetry Breaking.
This would be accomplished in practice through the physical
production of one or more Higgs boson particles. Undoubtedly, the
next-to-most important discovery expected at the LHC is the finding
of supersymmetric particles.

Actually, the discovery of Supersymmetry (SUSY)
(see\,\cite{Ferrara87} for a comprehensive review) is intimately
connected to the structure of the Higgs mechanism. In fact,
unearthing supersymmetric particles would be strong evidence that
Higgs bosons (in plural) should be around the corner. The opposite,
however, is not necessarily true, but if a light Higgs boson of,
say, $130\,\GeV$ would be found at the LHC, the hopes for SUSY
physics would stay high and we would immediately felt encouraged to
search for more Higgs bosons and potential supersymmetric particles.
It is well-known that a light Higgs boson ($m_h<140\,\GeV$) is a
trademark prediction, if not of SUSY in general, at least of the
Minimal Supersymmetric Standard Model (MSSM) in particular, which is
after all the canonical scenario for low-energy SUSY
phenomenology\,\cite{\MSSM}.

If SUSY is realized at the $\TeV$ scale (usually taken as the
characteristic energy scale to explain the naturalness problem of
the SM\,\cite{Haber:1985rc}), one expects that a few (or even a
bunch of) supersymmetric particles of the MSSM spectrum should be
well reachable at the LHC. However, the tagging of heavy new
particles is not an easy task because of the many decay modes
available, most of them carrying invisible neutral species (some of
them also of genuine SUSY origin, like sneutrinos and neutralinos)
and, therefore, leading to missing energy events -- usually hard to
interpret. For this reason, one expects to get a complementary clue
to the underlying SUSY dynamics from the short-distance quantum
corrections on more conventional processes. If these supersymmetric
quantum effects can be measured, they can be a solid handle to the
properties of the new physics. The idea has been known for a long
time and has been applied to the familiar physics of the $\PW$ and
$\PZ$ gauge bosons, see
e.g.\,\cite{Grifols:1984xs,Garcia:1993sb,Garcia:1994wu,Garcia:1994wv}.
Here we wish to apply this method to the realm of rare processes,
namely processes with conventional initial and final states which,
although not strictly forbidden, turn out to be highly suppressed
within the SM context. Among them, we have the fruitful
Flavor-Changing Neutral-Current (FCNC) processes.

The study of the flavor-changing interactions, in particular the
FCNC processes, has been a very active field of research for about
forty years, namely as of the glorious times when Glashow,
Iliopoulos and Maiani (GIM) successfully proposed the existence of a
fourth species of quark, the c-quark, to suppress to an acceptable
level the strangeness-changing neutral-current effects in rare
processes (e.g. $\HepProcess{\PKlong \HepTo \APmuon \Pmuon}$) that
otherwise would proceed at the tree-level, and similarly to further
suppress the one-loop contributions in e.g. the $\PKzero-\APK$
system. Indeed, it was the experimental evidence that the FCNC
processes seemed to be extremely inhibited in nature (actually
forbidden at the tree-level and highly suppressed at the one-loop
level) the main motivation for the aforementioned GIM
mechanism\,\cite{\GIM}, nowadays embedded in a natural way into the
current formulation of the SM -- essentially into the unitarity of
the CKM matrix. It is remarkable, however, that the degree of
suppression at one-loop order can vary from one process to another
in a dramatic manner. For instance, in the $\Pbottom$-quark sector
the radiative \PB-meson decay has a branching ratio
$\mathcal{B}(\HepProcess{\Pbottom \HepTo \Pstrange \Pphoton}) \simeq
3\times 10^{-4}$ which, although small, it has been measured
experimentally\,\cite{Yao:2006px} with quite some accuracy and it is
used in practice to constrain models of new physics. In contrast,
the FCNC top quark decay $\HepProcess{\Ptop \HepTo \Pcharm \Pgluon}$
becomes radically inhibited in the SM,
$\mathcal{B}(\HepProcess{\Ptop \HepTo \Pcharm \Pg}) \sim 10^{-11}$,
namely down to limits far below ever being possibly
observed\,\cite{\LE}. Amazingly, the top quark decay into the SM
Higgs boson is even more unlikely:
$\mathcal{B}(\HepProcess{\Ptop\HepTo\Pcharm\PHiggs})\sim 10^{-14}$\,
\cite{Eilam:1990zc,Mele:1998ag}. In all these cases, it is their
highly ``expected unobservability'' what provides the natural
``signature'' for potentially unraveling new physics out of their
study. In fact, the huge GIM suppression in some rare processes
within the SM can be significantly softened if one accounts for
possible SUSY virtual contributions. For example, in the case of the
extremely rare top quark decay into the SM Higgs boson one can show
that if $\PHiggslight^0$ is the lightest CP-even Higgs boson in the
MSSM\,\cite{Hunter}, then $\mathcal{B}(\HepProcess{\Ptop\HepTo
\Pcharm\,\PHiggslight^0})$ can be enhanced $10^{10}$ times as
compared to the SM mode and, thus, bring it to the observable level
$\sim 10^{-4}$ \,\cite{Guasch:1999jp,Guasch:1997kc,Yang:1993rb}.
Similar results hold for the Higgs boson decay modes into heavy
quarks, see e.g.
\cite{Curiel:2002pf,Demir:2003bv,Curiel:2003uk,Bejar:2004rz,Arhrib:2006vy}.
Actually, not only SUSY can help here; other alternative extensions
of the SM, among them the general Two-Higgs-Doublet Model
(2HDM)\,\cite{Hunter}, predict in some cases an enhanced, and often
distinctive, FCNC
phenomenology\,\cite{\Santi,\Arhrib,Bejar:2005kv,\fer}. Put another
way: by finding experimental evidence of non-standard FCNC processes
we can not only enlighten the existence of physics beyond the SM
but, in favorable conditions, we can even tell the kind of new
physics hiding right there\,\footnote{For a review, see e.g.
\cite{Bejar:2001sj} and \cite{Bejar:2006ww}. See also the
interesting flavor mixing studies
\cite{Cao:2007dk,Cao:2006xb,DiazCruz:2001gf}.}.

In this paper we wish to further explore the FCNC physics of the top
quark, but in this case we focus on the production of single-top
quark final states $\Ptop\APcharm$ or $\APtop\Pcharm$ through gluon
fusion ($\Pgluon\Pgluon$) in $\Pproton\Pproton$ collisions to take
place at the LHC. We denote it by $\processgg$.  While it is true
that this process is possible within the strict SM, it proceeds
through (GIM-supressed) charged-current interactions. We have found
the following cross-section for this process at one-loop level (see
Section \ref{sec:results}):
\begin{equation}\label{SMtc}
\sigma (\processgg)_{\rm SM}=8.46\times 10^{-8}\,\text{pb}\,.
\end{equation}
Obviously, it is so tiny that it amounts to less than one event in
the entire lifetime of the LHC! \ So it is pretty clear that if this
kind of FCNC-generated single top quark signatures would ever be
detected at the LHC, if only at a level of a few dozen crystal-clear
events, then the presence of new physics could perhaps be the only
valid explanation for them. We see that the situation with this
production process is very similar to the rare top quark decay modes
mentioned above; in both cases it is the FCNC physics of the top
quark that provides the extreme suppression within the SM. However,
it should be clear that the top quark final states in (\ref{SMtc})
are a particular class of events within the large variety of single
top quark processes available in hadron
colliders\,\cite{Stelzer:1998ni,Sullivan:2004ie,Cao:2004ap,Cao:2005pq}.

But this is not the only challenge. The LHC, with all its ability to
dig deep beneath the physics of the top quark, could perhaps be
sensitive to the class of single top quark final states associated
to FCNC processes, $\processgg$, provided of course the underlying
mechanism could be sufficiently enhanced by some form of new physics
capable to boost its cross-section up to $\sim$pb level. In this
study, we will show that the necessary enhancement (which amounts to
a factor of roughly $10^7$ in the total cross-section) could just
come from the world of the supersymmetric interactions in the
general MSSM.

Interestingly enough, let us remark that for the FCNC process under
consideration, $\processgg$, there is no significant competition
between the MSSM and the general 2HDM because there is no
enhancement to speak of from the latter. This is in contrast to the
situation with the rare top quark decays mentioned above, where the
2HDM contributions are non-negligible as compared to the MSSM ones.
Moreover, the direct production mechanism $\processgg$ is
substantially more efficient (typically a factor of $100$) than the
production and subsequent FCNC decay of the heavy Higgs bosons
($\HepProcess{\PHiggspszero\,, \PHiggsheavyzero \HepTo \Ptop\APcharm
 + \APtop \Pcharm}$)\,\cite{Bejar:2005kv}. In
this sense the discovery of a bunch of well-identified ${\rm
\Ptop\APcharm}$ and/or ${\rm \APtop\Pcharm}$ events could be strong
evidence, not only of new physics, but perhaps of SUSY itself. Some
of these features were already emphasized in
Ref.\,\cite{Guasch:2006hf}, where it was presented a first
self-consistent study of this subject (see also \cite{JS:RC05}).
These references, however, reported on the computation of the
SUSY-QCD effects only. Other studies can be found in
\cite{Eilam:2006rb,Liu:2004bb} under different sets of assumptions.
In our case we will continue within the general approach initiated
in \cite{Guasch:1999jp}, and continued in\,\cite{Guasch:2006hf}. It
means that the flavor-mixing coefficients $\delta_{ij}$ will be
allowed only in the purely left-handed part of the $6\times 6$
sfermion mass matrices in flavor-chirality space, as it is indeed
suggested by standard renormalization group (RG)
arguments\,\cite{Duncan:1983iq,Nilles:1984ex}. Within this well
motivated setup we provide here a full treatment of the SUSY-EW
effects and combine them with the SUSY-QCD ones\cite{Guasch:2006hf}
within the general framework of the MSSM. We wish to remark that, in
contradistinction to the aforesaid studies by other authors, we
present our MSSM calculation of $\processgg$ in combination with the
corresponding MSSM effects on the low-energy $\bsg$ decay and,
therefore, we extract the single top quark FCNC results only in the
region of parameter space compatible with the experimental bounds on
the radiative B-meson decays. This procedure is, in our opinion, a
self-consistent approach to the computation of the FCNC single top
quark signal under study.

The paper is organized as follows. Section\,\ref{sec:formalism} is
devoted to the general formalism for the FCNC processes in the MSSM.
In section  \ref{sec:top} we summarize the details of our
calculation of $\sigma(\processgg)$ in this framework. The full
numerical analysis is presented in section \ref{sec:results},
leaving section \ref{sec:discussion} to discuss the results and
deliver our conclusions.

\section{Formalism: FCNC interactions in the MSSM}
\label{sec:formalism} Apart from the conventional charged-current
flavor changing interactions in the SM, the FCNC processes in the
MSSM are driven by explicit intergenerational mixing terms arising
from the mass sector of the squarks. For a brief review of this
topic, our starting point shall be to specify the form of the
superpotential, which is the crucial piece of any SUSY theory of
particle interactions. In our case we will consider the  MSSM with
arbitrary soft-SUSY-breaking terms. The most general
gauge-invariant form of the superpotential can be cast in terms of
chiral superfields (denoted by a hat) as follows\,\cite{\MSSM}:
\begin{eqnarray}
W_{MSSM} &=& \epsilon_{rs} \left[y_l \hat{H}_1^r \hat{L}^s \hat{E}
+ y_d \hat{H}_1^r \hat{Q}^s \hat{D}  + y_u \hat{H}_2^s \hat{Q}^r
\hat{U} - \mu \hat{H}_1^r \hat{H}_2^s \right] \label{eq:wmssm}.
\end{eqnarray}
Indices $r,s=1,2$ refer to the components of the $SU(2)_L$ doublets,
which are combined in a gauge-invariant form through $\epsilon_{rs}$
(with $\epsilon_{12}=-\epsilon_{21}=1$). The set of parameters
$y_l$, $y_d$ and $y_u$ constitute Yukawa coupling $3\times 3$
matrices in generation space. Although explicit generation labels
have been suppressed here, they will be introduced at due time. Let
us notice that in more general SUSY theories there are additional
pieces of the superpotential inducing violation of baryon or lepton
number, but in the MSSM they are set to zero because one assumes
that the R-parity symmetry holds.

\noindent We also need to settle the piece of the soft
SUSY-breaking Lagrangian that takes part in the squark mass
matrix:
\begin{eqnarray}
\lag_{soft} = &-& M_{\tilde{Q}} \tilde{Q}^{*}\tilde{Q} -
M_{\tilde{U}}^2
\tilde{U}^{*}\tilde{U} - M_{\tilde{D}}^2 \tilde{D}^*\tilde{D} \nonumber \\ \nonumber \\
&-& \frac{g}{\sqrt{2}\,M_W}\epsilon_{rs} \left[ \frac{m_d
A_d}{\cos\beta}\, H_1^r \tilde{Q}^s \tilde{D} - \frac{m_u
A_u}{\sin\beta}\, H_2^r \tilde{Q}^s \tilde{U} \right] + h.c.
\label{soft}.
\end{eqnarray}
In this expression, $Q$ stands for the $SU(2)_L$ quark doublets,
while $U,D$ denote the corresponding singlets. Let us recall that
each of the above mass and trilinear coupling parameters carries a
$3\times 3$ matrix structure in the flavor space, although we shall
not keep track of it explicitly.

We can now collect the different pieces contributing to the general
form of the squark mass matrix, which come either from the explicit
mass terms in the soft-SUSY-breaking Lagrangian (\ref{soft}) or from
the couplings triggered by the superpotential (\ref{eq:wmssm}) after
spontaneous symmetry breaking (SSB) of the EW symmetry. If we
arrange all such terms in a $2$-dimensional left-right chirality
space, we are left with the following mass matrix:
\begin{equation}
\mathcal{M}_{\tilde{q}}^2  = \left(
\begin{array}{cc}
M_{\tilde{Q}\,L}^2 + m_q^2 + \cos 2\beta (T_3^{q_L} - Q_q\, \sin^2
\theta_W \,)\, M_Z^2 &
m_q \, M_{LR}^q \\
m_q \, M_{LR}^q & M_{\tilde{Q}\,R}^2 + m_q^2 + \cos 2\beta\, Q_q\,
\sin^2 \theta_W \, M_Z^2 \end{array} \right) \label{eq:mass1},
\end{equation}
where in the off-diagonal mass terms we have defined $M_{LR}^{u} =
A_u - \mu \cot \beta$ and $M_{LR}^{d}=A_d - \mu \tan \beta$. As
usual, $\tan \beta =<H_2^0>/<H_1^0>\equiv {v_2}/{v_1}$, with
$v_1^2+v_2^2=\,G_F^{-1}/\sqrt{2}$, defines the ratio of the vacuum
expectation values of the two Higgs doublets giving masses to the up
and down quarks respectively, while $T_3$ stands for the $3$th
component of the weak isospin of the left-handed quark $q_L$, and
$Q_q$ denotes its charge. The non-diagonal structure of
(\ref{eq:mass1}) in the chirality basis requires its diagonalization
in order to obtain the physical mass-eigenstates $\tilde{q}_a$ in
terms of the electroweak (EW) squark eigenstates $\tilde{q}_a^{'}$
with well-defined $SU(2)_L\times U(1)_Y$ quantum numbers. If
$R^{(q)}$ denotes the matrix rotating the $q$th flavor, we can
diagonalize the mass matrix as follows: $R^{(q)\,\dagger} \,
\mathcal{M}^2_{\tilde{q}} R^{(q)} = {\rm diag}
(m^2_{\tilde{q}_1},m^2_{\tilde{q}_2})$. Notice that each matrix
elements in Eq~(\ref{eq:mass1}) is proportional to the unity matrix
$1_{3\times 3}$ in the flavor space.  It is worth realizing,
however, that such a trivial flavor structure for the mass matrix
does not provide the most general realization of the squark mass
sector. Indeed, in the MSSM we have two fundamental sources of
flavor violation. One of them just mimics the SM one, namely it
consists of the flavor mixing among up- and down-like squarks
triggered by the charged-current interactions induced by the charged
gauge bosons, the charged Higgs bosons and the charginos. The second
one is qualitatively new and is caused by the so-called
\emph{misalignment} between the rotation matrices that diagonalize
the quark and squark sectors or, in other words, the fact that the
squark mass matrices in general need not diagonalize with the same
matrices as the quark mass matrices\,\cite{\duncan,\BurasGabbiani}.
This is reflected in the existence of the gaugino-fermion-sfermion
interactions mediated by gluinos ($\PSgluino^a;\ a=1,2,...,8$) and
neutralinos ($\chi_{\alpha}^0;\ \alpha=1,2,..,4$). Consider e.g. the
gluino-quark-squark interactions
\begin{eqnarray}
\lag_{{\PSgluino}q\sq} &=& -i\sqrt{2} \, g_s \bar{{\PSgluino}}^a
\left\{\,{\su}^*_{L_i}\,V(u)_{ij}\,(T^a)
u_{L_j}+{\sd}^*_{L_i}\,V(d)_{ij}\,(T^a) d_{L_j}\right\} + h.c.
\label{eq:gluino}\,,
\end{eqnarray}
with
\begin{equation}\label{UV}
V(u)\equiv B^{\dagger}(\su_L)A(u_L)\,,\ \ \ \ \ \ \ \ V(d)\equiv
B^{\dagger}(\sd_L)\,.
\end{equation}
Here $i,j$ are generation indices, $T^a$ are the $SU(3)_c$
generators, and $A(u_L),B(\su_L) $ and $B(\sd_L)$ are rotation
matrices in generation space which relate the electroweak and the
mass-eigenstates; e.g. $ A(u_L)$ rotates up-quarks and $B(\su_L)$
rotates up-squarks, etc. Notice that in the down sector we only need
to rotate squarks through $B(\sd_L)$ because after SSB of the gauge
symmetry the down quark matrix is already diagonal. This follows
from the fermion mass matrix structure that emerges from the
superpotential (\ref{eq:wmssm}) in generation space after the
Higgs bosons acquire VEV's and spontaneously break the EW symmetry.
Let us consider only the quark sector,
\begin{equation}\label{Yuk}
\lag_{m_q}= - \bar{q}_{L_{i}}  \left(\begin{array}{c}
                         0 \\
                         v_1/\sqrt{2} \\
                       \end{array}\right)\,(y_d)_{ij}\ {d}_{R_j}
 -  \bar{q_L}_i\, \left(\begin{array}{c}
                         v_2/\sqrt{2} \\
                         0 \\
\end{array}\right)\, (y_u)_{ij}\ {u_R}_j +\,h.c.
\end{equation}
We can rotate ${q}_{L_{i}}$ and $ {d}_{R_j}$ in generation space
until the mass matrix for down quarks, $(v_1/\sqrt{2})\,(y_d)_{ij}$,
becomes diagonal, but then the mass matrix for up quarks,
$(v_2/\sqrt{2})\,(y_u)_{ij}$, will in general be non-diagonal
because ${q}_{L_{i}}$ was already rotated. By inspecting the charged
current interaction of quarks, this immediately implies that the
ordinary CKM matrix is just $U_{\rm CKM}=A^{\dagger}(u_L)$.
Similarly, from the charged current for squarks we read off the
corresponding CKM matrix in the squark sector: $U_{\rm
SCKM}=B^{\dagger}(\su_L)B(\sd_L)$. Therefore one finds a relation
between the CKM and SCKM matrices:
\begin{equation}\label{CKMs}
    U_{\rm SCKM}=V(u)\,U_{\rm CKM}\,V(d)^{\dagger}\,.
\end{equation}
As a result, in the MSSM we need three unitary matrices to
parametrize the flavor changing interactions, one is the ordinary
CKM matrix and the other two are associated to the new FCNC
gaugino-quark-squark couplings (\ref{eq:gluino}). Neglecting this
second source of flavor changing interactions (i.e. assuming that
the matrices $V(u)$ and $V(d)$ are unity in flavor space) would be
tantamount to assume that quarks and squarks diagonalize
simultaneously, i.e. $U_{\rm SCKM}=U_{\rm CKM}$. This is the
super-CKM basis approach to the FCNC processes; it assumes that
these processes appear at one-loop only through the charged current
interactions (from $W^{\pm}_{\mu}$, charged Higgs bosons $H^{\pm}$
and charginos $\chi^{\pm}$) and with the same mixing matrix elements
as in the Standard Model CKM matrix. However, in general we expect
that the two sources of FCNC should be active in the MSSM and we
will take them both into account in our calculation.

These observations turn out to be crucial for the discussion of the
flavor-changing processes in the MSSM because it means that we can
extend the simple $2\times 2$  squark mass matrices in chiral space
into $6\times 6$ mass matrices in (flavor)$\otimes$(chiral) space.
We shall comment below on how to parametrize the flavor mixing
terms. For the moment we note that, due to the aforementioned flavor
mixing, the squark mass matrix diagonalization process must be
extended as follows:
\begin{eqnarray}
\sq_{a }^{\prime } &=&\sum_{b=1}^{6}R_{ab}^{(q)}\sq_{b}\,, \ \ \ \
(a=1,2,...,6) \nonumber  \label{eq:definicioR6gen} \\
R^{(q)\dagger }\mathcal{M}_{\sq}^{2}\,R^{(q)} &=&\mathrm{diag%
}\{m_{\sq_{1}}^{2},\ldots ,m_{\sq_{6}}^{2}\}\,\,\,\,\ \ (q\equiv
u,\,d)\,\,,
\end{eqnarray}
where $\mathcal{M}_{(\su,\sd)}^{2}$ are the $6\times 6$ square mass
matrices for squarks in the EW basis ($\sq_{\alpha }^{\prime }$),
the eigenvalues being denoted $m_{\sq_{a}}^{2}$. Indices run now
over $6$-dimensional space vectors with suitable identifications.
For example, for up-type squarks
$a =1,2,3,\ldots ,6\equiv \su_{L},\su_{R},\tilde{c}_{L},\ldots ,\stopp%
_{R}$, and a similar assignment for down-type squarks. Furthermore,
let us notice that the $SU(2)_L$ gauge invariance of the MSSM
Lagrangian imposes certain restrictions over the up-squark and
down-squark $6\times 6$ soft-SUSY breaking mass matrices,
specifically in their LL blocks, as follows:
\begin{eqnarray}
\left( M^2_{\tilde{U}} \right)_{LL} &=&
K\,\left(M^2_{\tilde{D}}\right)_{LL} K^\dagger \label{eq:relsu2},
\end{eqnarray}
where $K$ stands for the CKM matrix (previously denoted $U_{CKM}$ for convenience).
It is thus clear, in particular, that both squark matrices cannot be simultaneously
diagonal (unless they are proportional to the identity) and,
therefore, they cannot be simultaneously diagonal with the up-like
and down-like quark mass matrices either. This is again a reflect of
the \emph{misalignment} effect between the mass matrices of quarks
and squarks.

Despite what we have just argued above, within the context of Grand
Unified Theories (GUT's), one usually assumes that the parameters
should be aligned at the characteristic high energy scale $M_X\sim
10^{16}\,GeV$ of these theories (that is to say, the quark and
squark mass matrices should diagonalize simultaneously at $M_X$).
But even within such theoretically-motivated scenario, it can be
shown that the renormalization group running of these parameters
down to the EW scale would again destroy the primeval aligned
configuration\,\cite{\duncan,Nilles:1984ex}. It is therefore wiser
to take the misalignment into account right from the start in the
calculation. The most common way to parametrize it is by defining
the following dimensionless quantities, $\delta^{AB}_{ij}$, being
$A,B = 1,2$ the chirality indices and $i,j = 1,2,3$ the flavor ones,
in such a way that we can set the non-diagonal squark mass matrix
elements to be:
\begin{eqnarray}
\left(M^{2\;AB}_{ij}\right) &=& \delta^{AB}_{ij} \tilde{m}_i^A
\tilde{m}_j^B \ \ \ \ (i\neq j)\label{eq:relacio},
\end{eqnarray}
where $\tilde{m}^A_i$ stands for the soft-SUSY breaking parameter of
a given chirality and flavor. (No sum over repeated indices here.)
It is very common to set all the mass parameters $\tilde{m}^A_i$
equal to a generic SUSY scale $M_{SUSY}$.

As far as we are dealing with FCNC processes involving the top
quark, the most relevant mixing parameters are those ones relating
the heavy up-like flavors among themselves, thus essentially
$\Ptop-\Pcharm$ transitions parametrized by  $\delta_{23}(u)$ above.
In close relation to them  we have the $\Pbottom-\Pstrange$
transitions controlled by the parameter $\delta_{23}(d)$. Only these
mixing parameters are expected to be large in GUT's and, moreover,
they are not significantly constrained by phenomenological
considerations. The experimental bounds on the various mixing
parameters are derived from the absence of low-energy FCNC
processes, which mainly involve the first and second generations.
For instance, the measurements of the mass splitting in $\PK -
\APKzero$ and $\PD - \APDzero$ phenomena \cite{\Pokorsky}.

Regarding heavy flavors, the phenomenological constraints come from
the branching ratio of the radiative B-meson decay
$\mathcal{B}_{exp} (\bsg)$  and also from the mass splitting in $\PB
- \APBzero$ mixing effects. Clearly, such two processes can only be
sensitive to the down-like heavy-flavor mixing parameter,
$\delta_{23}(d)$. However, they can also provide information on the
allowed values for $\delta_{23}(u)$ since both up and down-like
flavor-mixing parameters must necessarily be related through the
$SU(2)_L$ symmetry (\ref{eq:relsu2}). As advertised, in our
framework we limit ourselves to consider flavor mixings only in the
LL-block of the squark mass-matrices, the only ones which are
well-motivated by RG arguments. Thus, the relevant piece in our
calculation will be the LL sector of the up-type squark $6\times 6$
mass matrix, which can be rewritten in the following manner:
\begin{equation}
\left( M^2_{\tilde{u}}\right)_{LL} = M^2_{SUSY} \left(\begin{array}{ccc} 1 & 0 & 0 \\
0 & 1& \delta_{23}(u) \\ 0 & \delta_{23}(u) & 1
\end{array}\right)_{LL} \label{eq:mass2}\,.
\end{equation}
Similarly for $M^2_{\tilde{d}}$ with $\delta_{23}(u)\to
\delta_{23}(d)$. Squark mass-eigenstates follow from diagonalization
of these matrices through Eq.~(\ref{eq:definicioR6gen}). The
mass-eigenstates of (\ref{eq:mass1}) are recovered by setting the
mixing parameters $\delta$ to zero, as could be expected.

Once we have discussed where the flavor-mixing source is rooted in
the MSSM, we must now trace back its role at the Lagrangian level
\cite{Rosiek:1995kg}. The misalignment between the diagonalization
matrices in the quark and squark mass sectors triggers the presence
of couplings of the guise gluino-quark-squark (in the SUSY-QCD part)
and neutralino-quark-squark (in the SUSY-EW one) that allow the
interaction of quarks having the same charge but belonging to
different generations. At the $1$-loop level it is also possible to
have this kind of flavor-changing interactions mediated by the
ordinary SM charged currents, but in the $R$-odd part of the MSSM we
also have the chargino-up-quark-down-squark interactions and the
charged Higgs-up-quark-down-squark vertices. In all such cases the
SUSY nature of the couplings allow the resulting process to bypass
the SM GIM mechanism and provide non-suppressed FCNC events (the
charged Higgs piece is an exception, as we shall see). The
importance of such effect is correlated with the choice of the MSSM
parameters, in particular those specifying the soft-supersymmetry
breaking and, of course, the explicit intergenerational mixing
$\delta_{ij}^{AB}$, which are the most relevant ones for the
flavor-changing dynamics in the MSSM. For the SUSY-QCD coupling one
must work out the supersymmetrized gauge interaction piece:
\begin{eqnarray}
\lag_{\tilde{\lambda}\psi\tilde{\psi}} &=& -i\sqrt{2} \, g_s
\tilde{\psi}^*_k \tilde{\lambda}^a (T^a)_{kl} \psi_l + h.c.
\label{eq:gauge},
\end{eqnarray}
where $T^a$ are the gauge group generators, the indices ${k,l}$
denote the corresponding gauge quantum numbers (color, weak isospin)
of the interacting particles and $\tilde{\lambda}^a$ stands for a
generic gaugino field. To extract the FCNC vertices one must include
the generation indices in these interactions. For the particular
case of the gluino-mediated interactions, this was done in
(\ref{eq:gluino}). For the practical calculations we will use the
extended $6\times 6$ diagonalization matrices defined in
(\ref{eq:definicioR6gen}). Therefore, by plugging the squark
mass-eigenstates in this expression we can rephrase the result in
the mass-eigenstate basis and in terms of four-component Dirac
spinors (for both the gluino and quarks). In the up quark-squark
sector we get
\begin{eqnarray}
\lag_{\PSgluino\,u\,\su} &=& -i\,\sqrt{2} g_s \sum_{a = 1}^{6}
\sum_{b = 1}^{3} \, \tilde{u}^*_{a} \left( R^{(q)\,*}_{a\,b}
\,\bar{\PSgluino}\, P_L - R^{(q)\,*}_{a\,(b+3)}\,\bar{\PSgluino} P_R
\right) u_b + h.c \label{eq:glu}\,,
\end{eqnarray}
and similarly for the down quark-squark sector. Here we have omitted
color indices for gluinos, quarks and squarks.  Notice that while the
sum over index $a$ runs over the whole flavor$\,\otimes\,$chirality
space, index $b$ runs only over generations because we are already
using the standard projectors $P_{L,R} = (1/2)(1\mp \gamma_5)$ to
set the chirality of the quarks.

A similar analysis can be performed to obtain the corresponding
Lagrangians describing the flavor-changing interactions in the
SUSY-EW sector. The calculations are slightly more involved since
such terms arise from the combination of the SUSY-gauge piece
(\ref{eq:gauge}) together with the higgsino-quark-squark Yukawa
couplings dictated by the superpotential (cf. Eq.~(\ref{eq:wmssm})).
Moreover, because of the EW symmetry breaking, the higgsinos and
gauginos mix together to give the final physical eigenstates, the
neutralinos $\tilde{\chi}^0_\alpha\ (\alpha=1,2,...,4)$ and
charginos $\tilde{\chi}^\pm_\beta\ (\beta=1,2)$. We shall quote here
the final result for such interaction Lagrangians (a detailed
derivation can be found in \cite{\haber} and references therein).
For the case of the neutralinos, we get:
\begin{eqnarray}
\lag_{\PSneutralino\Pup\PSup} &=& -i\,\sum_{\alpha=1}^4\,\sum_{a=1}^6\,\sum_{b=1}^3\, \tilde{u}_a^* \, \bar{\chi^0_\alpha}\,\Bigg[\frac{g}{\sqrt{2}}\,R^{(u)\,*}_{ab}\left(\frac{N_{1\alpha}}{3}\,\tan\theta_W + N_{\alpha\,2}\right)\,P_L + y_u\,R^{(u)\,*}_{a(b+3)}N_{\alpha\,4}\,P_L + \nonumber \\
&& +y_u\,R_{ab}^{(u)\,*}\,N_{\alpha\,4}\,P_R -
\frac{4\,g}{3\,\sqrt{2}}\,\tan\theta_W \,
R^{(u)*}_{a(b+3)}\,N_{\alpha\,1}^* P_R
 \Bigg]\, u_b + h.c.
\label{eq:neut}\,,
\end{eqnarray}
where $g$ is the weak  $SU(2)_L$ gauge coupling constant. A few
words about notation: index $\alpha$ is running over the four
neutralino states, $N_{\alpha \beta}$ being the diagonalization
matrix that provides the neutralino mass-eigenstates, $N^{*}
M_{\tilde{\chi}_0} N^{\dagger} = diag ({m_{\tilde{\chi_0}}}_1,
{m_{\tilde{\chi_0}}}_2, {m_{\tilde{\chi_0}}}_3,
{m_{\tilde{\chi_0}}}_4)$, while $y_u$ stands for the corresponding
Yukawa coupling and $\theta_W$ is the weak mixing angle
($e=g\,\sin\theta_W$). Similarly the chargino-up-squark-down-quark
interaction Lagrangian can be cast in the following form:
\begin{eqnarray}
\lag_{\PSino\Pup\PSdown} &=&
-i\,\sum_{\beta=1}^2\,\sum_{a=1}^6\,\sum_{b=1}^3\,\sum_{c=1}^3
\tilde{d}_a^*\,\bar{\chi_\beta} \Bigg[g\,R_{ab}^{(d)}\, U_{\beta
1}\,P_L \nonumber \\ &&+ y_d\, R^{(d)}_{a\,(b+3)}\,U_{\beta
2}\,P_L - y_u\,R^{(d)}_{ab}\,V_{\beta 2}^*\,P_R \Bigg]\,
K_{bc}^*\,u_c + h.c.
\label{eq:cha}.
\end{eqnarray}

\noindent This time the standard CKM matrix also needs to be taken
into account because of the charged-current mixing between up-like
squarks with down-like quarks. Again, $U,V$ refer to the
diagonalization matrices of the chargino mass, such that $U^{*}
M_{\PScharginopm}V^{\dagger} = diag(m_{\chi^+},m_{\chi^-})$.
\section{Single top-quark production through FCNC processes in the MSSM: computation procedure}
\label{sec:top}

In the following we will concentrate on the analysis of the single
top-quark production by direct supersymmetric flavor-changing
interactions at the LHC, namely the processes leading to $\Ptop
\APcharm$ or $\APtop \Pcharm$ final states. The leading mechanism is
the gluon fusion channel: $\processgg$ (see Section
\ref{sec:results} for a full list of Feynman diagrams). It should be
clear that $\sigma(\processgg) = 2\,\sigma(\HepProcess{\Pproton
\Pproton(\Pgluon\Pgluon) \HepTo \Ptop \APcharm})$. It was already
shown in \cite{\Liu} that the $\Pgluon\Pgluon$ partonic channel
largely dominates over the $q\bar{q}$ one at the LHC. Although there
are previous studies of this process in the literature within the
MSSM and adopting different
approximations\,\cite{\main,\Liu,\eilamsis}, a closer look is highly
desirable from our point of view. This is so because the kind of
simplified assumptions made in some of the previous analyses do not
shed sufficient light on the possibility that this process could be
sufficiently enhanced in the MSSM as to be considered realistically
at the LHC. We will comment on the differences among these
approaches later on. In the present paper we carry out our
calculation within the framework of \cite{\main,JS:RC05}, which was
first delineated in \cite{Guasch:1999jp}.

Throughout the present work we have made use of the standard
algebraic and numerical packages \textit{Feynarts},
\textit{FormCalc} and \textit{LoopTools} \cite{\FeynArts} for the
obtention of the Feynman diagrams, the analytical computation and
simplification of the scattering amplitudes and the numerical
evaluation of the cross section (up to the partonic level). Notice,
however, that we also need to address the computation of the total
hadronic cross section in order to account for the physical process,
a $\Pproton\Pproton$ collision, to take place at the LHC. To this
aim we have made use of the program \textit{HadCalc} \cite{\rau}
\footnote{The source code is available on request from the author.},
while several cross-checks have also been performed with other
independent codes implemented by us. Throughout our calculations we
have settled both renormalization and factorization scales at a
common value, chosen to be half of the production threshold $\mu_R =
\mu_F = \frac{1}{2} (m_c + m_t)$. Concerning the parton distribution
functions (PDF's) involved in the long-distance dynamics of the
hadronic process, we have included the recent CTEQ6AB data set
\cite{\cteq} provided by Les Houches Accord Parton Distribution
Functions Library (v.5.2) \cite{\lhapdf}.

The computation of  $\sigma(\processgg)$ in the MSSM is not
straightforward. It involves a number of subtleties that must be
carefully handled. To begin with, we must deal with the PDF of a
gluon, which exhibits a huge slope in the low-momentum region. To
that purpose, we have implemented a logarithmic mesh for the
integration over the partonic variables instead of the linear mesh
that is provided by default. We must obviously pay the price of
adding the corresponding jacobian piece to the original integral.
Furthermore, a double call to the integration subroutine (viz. the
\textit{Vegas} routine provided by the Cuba library\,\cite{\vegas})
has been implemented. The first call provides an adapted grid for
the second one, in such a way that the convergence is much faster.
As a result a good and reliable numerical accuracy is achieved
(meaning that the $\chi^2$ values always remain of order $1$). Last
but not least, a second non-trivial subtlety emerges from the fact
that one of the final states, the $\Pcharm$-quark, has a very small
mass when compared to the $\sqrt{S}$ value of the scattering
process. We are thus very close to a collinear-divergence regime.
Although there is no analytical divergence, the mass of the c-quark
is low enough to trigger instabilities in the code when integrating
over very small angles. We have carefully studied the problem and
have included a tiny angular cut (chosen to be such that $\sin^2
\theta< 0.03$) in order to avoid the aforementioned instability. We
have checked the dependence of the final calculation on the choice
of the angular cut (the total hadronic cross section can change
about a $15 \%$ when moving from $\sin^2 \theta< 0.01$ to $\sin^2
\theta< 0.1$), thus no dramatic changes occur when tuning the cut
within reasonable ranges.

In regard to the calculation of the amplitudes contributing to the
relevant process under consideration, $\processgg$, let us note that
the leading order is the $1$-loop level. This is a common feature
when studying FCNC processes in any renormalizable theory (due to
the lack of FCNC tree-level interactions). This implies that one
need not renormalize the bare parameters nor the Green's functions
as there are no explicit terms in the interaction Lagrangian where
to absorb the UV divergences. In other words, the overall amplitude
of the process should be already finite as soon as we add up all the
diagrams contributing to that process. In order to check the
finiteness of the resulting amplitude, we have made use of a
standard numerical procedure provided by \textit{FormCalc}.

In the following section we present our final numerical results. We
shall not provide here analytical details of the complicated
algebraic structures appearing in the calculation of the many
one-loop diagrams involved (see Figures 1-5). We have carried out
the computation in a fully automatic fashion by means of the
numerical and algebraic tools mentioned above, and of course we have
previously submitted our codes to many important tests and
non-trivial cross-checks of different nature.

The
calculation of $\sigma(\processgg)$ has been linked to the one-loop
calculation of $\mathcal{B}(\bsg)$ in the MSSM, so that after
enforcing this low-energy observable to stay within the experimental
bounds we have obtained the desired cross-section only in this
allowed region of the MSSM parameter space. Our computation of  $\mathcal{B}(\bsg)$ contains the complete leading
order (one-loop) MSSM computation including the flavor-violating
couplings. Specifically, we include the contributions to the high energy
operators from the SM ($W^\pm$ loops), SUSY-QCD (gluino loops) 
and SUSY-EW (chargino-neutralinos and Higgs boson
loops). The Wilson coefficient expressions have been taken from
Ref.\cite{Bobeth:1999ww}\,\footnote{Reference
\cite{Bobeth:1999ww} contains a computation of  $\mathcal{B}(\bsg)$
in the MSSM including some two-loop parts, but only the one-loop
contributions have been used for the present work.}, and they are
evolved using the leading order QCD renormalization group equations down
to the bottom mass scale. However, at certain stages of our work we use only
a part of these corrections, this will be clearly indicated in the text below.

\section{Numerical analysis}
\label{sec:results} To start with, let us present the calculation of
the cross-section for the process $\processgg$ within the context of
the SM. The Feynman diagrams describing the interaction at the
partonic level in the t'Hooft-Feynman gauge are shown in
Fig.~\ref{fig:dsm}. In this gauge, the covariant sum over the
polarization states of the gauge bosons yields the relation
$\sum_{\lambda} \epsilon^{*}(k,\lambda)\, \epsilon(k,\lambda) =
-g_{\mu \nu}$. Notice that in the present situation it is
unnecessary to introduce the Faddeev-Popov ghost-field contributions
since the current process involves only external gluon lines and it
thus suffices to restrict the above sum to the two physical degrees
of freedom carried by the gluons. This is straightforwardly done
within the framework of the standard computational tools of
Ref.\,\cite{\FeynArts} and allows us to get rid of the spurious
modes of the quantized gluon field. Also worth emphasizing is the
effect of the GIM suppression, which is associated to the SM
diagrams of Fig.~\ref{fig:dsm}. We take for instance a vertex
correction diagram driven by the exchange of a charged $\PW$ boson
with a pair of quark lines closing the loop, and then sum over
flavors. The result is a form factor of the type
\begin{eqnarray}
f \sim \frac{g^2}{16\pi^2} \, \sum_i \left(K^*_{ti}\,K_{ic} \right)
\, \left(\frac{m_i}{M_W}\right)^2 \label{eq:gim},
\end{eqnarray}
where $K_{ij}$ denote again matrix elements of the standard CKM
matrix, and $i$ is a flavor index that runs over the down-like quark
states $\Pdown, \Pstrange, \Pbottom$. We have also included a
standard numerical suppression factor from the one-loop integral.
The additional GIM suppression is of dynamical origin within the SM
and it stems from the unitarity of the CKM matrix. As a result the
overall behavior of the form factor amplitude goes like $\sim G_F
m_i^2$, where the $m_i$ correspond in this case to down-like quark
masses circulating in the loops. The cross-section
$\sigma_{t\bar{c}} \equiv \sigma (\HepProcess{\Pproton
\Pproton(\Pgluon\Pgluon) \HepTo \Ptop \APcharm})$ gets suppressed as
$\sigma_{t\bar{c}}\sim f^2$. It is thus not surprising that we
finally get the value early indicated in Eq.\, (\ref{SMtc}), which
amounts to about $10^{-4}$ fb. The order of magnitude of the
cross-section at the parton level roughly follows from naive power
counting and educated guess. Using (\ref{eq:gim}) we get:
\begin{equation}\label{ordmag}
\sigma (\HepProcess{\Pgluon\Pgluon\HepTo\Ptop\APcharm})\sim
\frac{|V_{bc}|^2}{s}\,\left(\frac{\alpha_s}{16\,\pi^2}\right)^2
\,(G_F\,m_b^2)^2\,,
\end{equation}
where we have included factors from the strong ($\alpha_s=g_s/4\pi$)
and weak ($G_F\sim g^2/M_W^2$) interactions. For the estimate we
include only the bottom quark contribution (with matrix elements
$V_{bc}\simeq 0.04$ and $V_{tb}\simeq 1$) as the other terms are
suppressed either by very small CKM matrix elements or very light
quark masses\,\footnote{This is similar to the kind of ansatz made
by Gaillard and Lee to predict the charm quark
mass\,\cite{Gaillard:1974hs}, except that here it is the bottom
quark that gives the dominant effect because the external quarks are
up-like.}. At the LHC energies the previous estimate provides the
cross-section within the ballpark of the exact result after
convoluting with the parton distribution functions. However, in
order to understand the dynamical mechanism of enhancement of the
SUSY interactions (see below) it will suffice to compare with the
partonic contribution (\ref{ordmag}).

Using the exact (numerically computed) result (\ref{SMtc}), we find
that this cross-section is literally invisible; even assuming a
total integrated luminosity of $1000$ \invfb\ it amounts to one
tenth of event during the whole lifetime of the LHC. This result
supports quite convincingly the idea that the eventual detection of
such kind of FCNC processes could give us important hints of some
form of physics beyond the SM.

\begin{center}
\EPSFIGURE[pt]{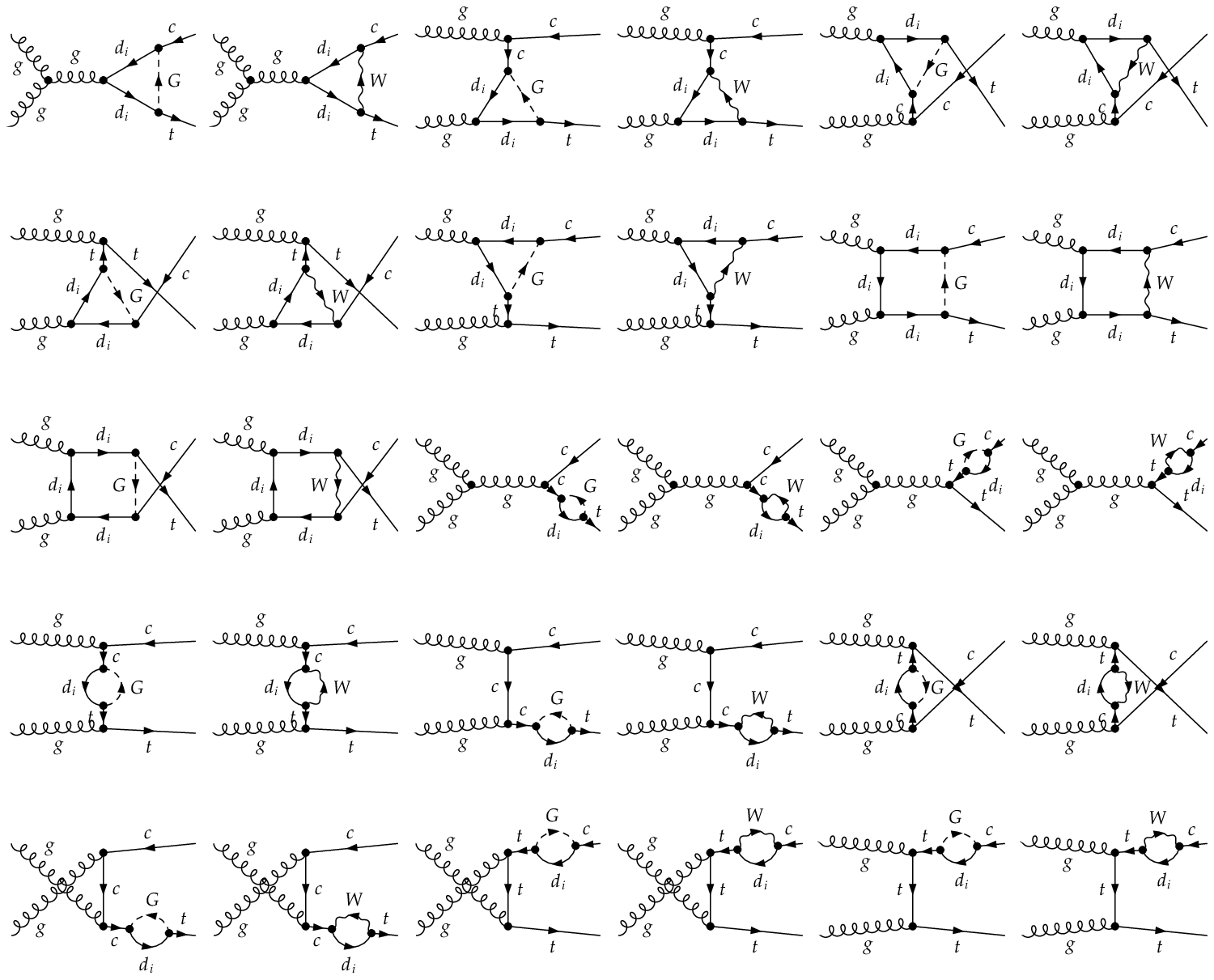}{Feynman diagrams corresponding to the
partonic process $\HepProcess{\Pgluon\Pgluon \HepTo \Ptop\APcharm}$
in the SM. There is a similar set of diagrams leading to the
$\APtop\Pcharm$ final states; this is implicitly understood here,
and also in the next four figures containing Feynman diagrams.
\label{fig:dsm}}
\end{center}

\begin{center}
\EPSFIGURE[pt]{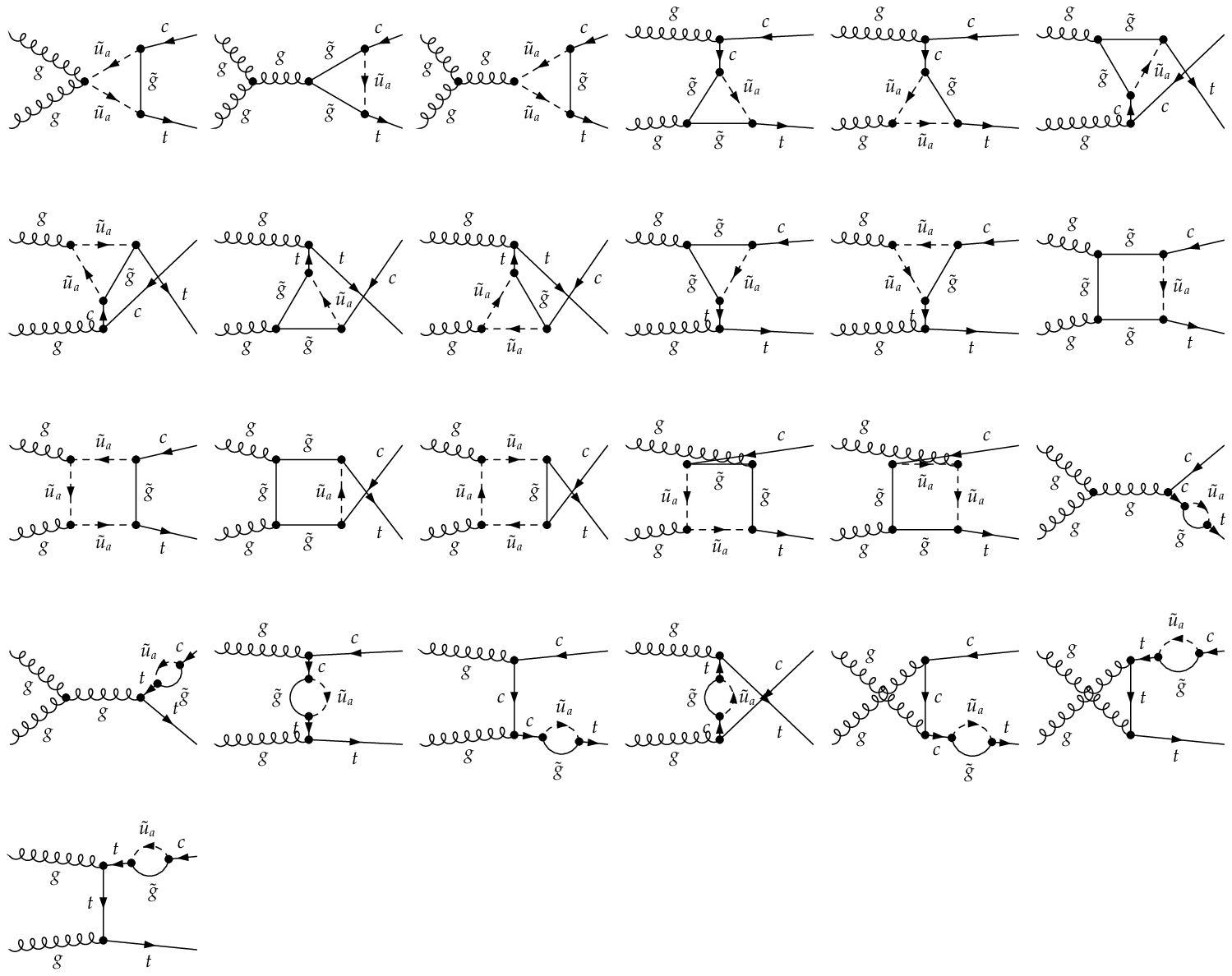}{Feynman diagrams involving gluino and squark
loops for the partonic process $\HepProcess{\Pgluon\Pgluon \HepTo
\Ptop\APcharm}$ in the MSSM. They constitute the SUSY-QCD
contribution to the single top quark production in the
MSSM.\label{fig:dglu}}
\end{center}

\begin{center}
\EPSFIGURE[pt]{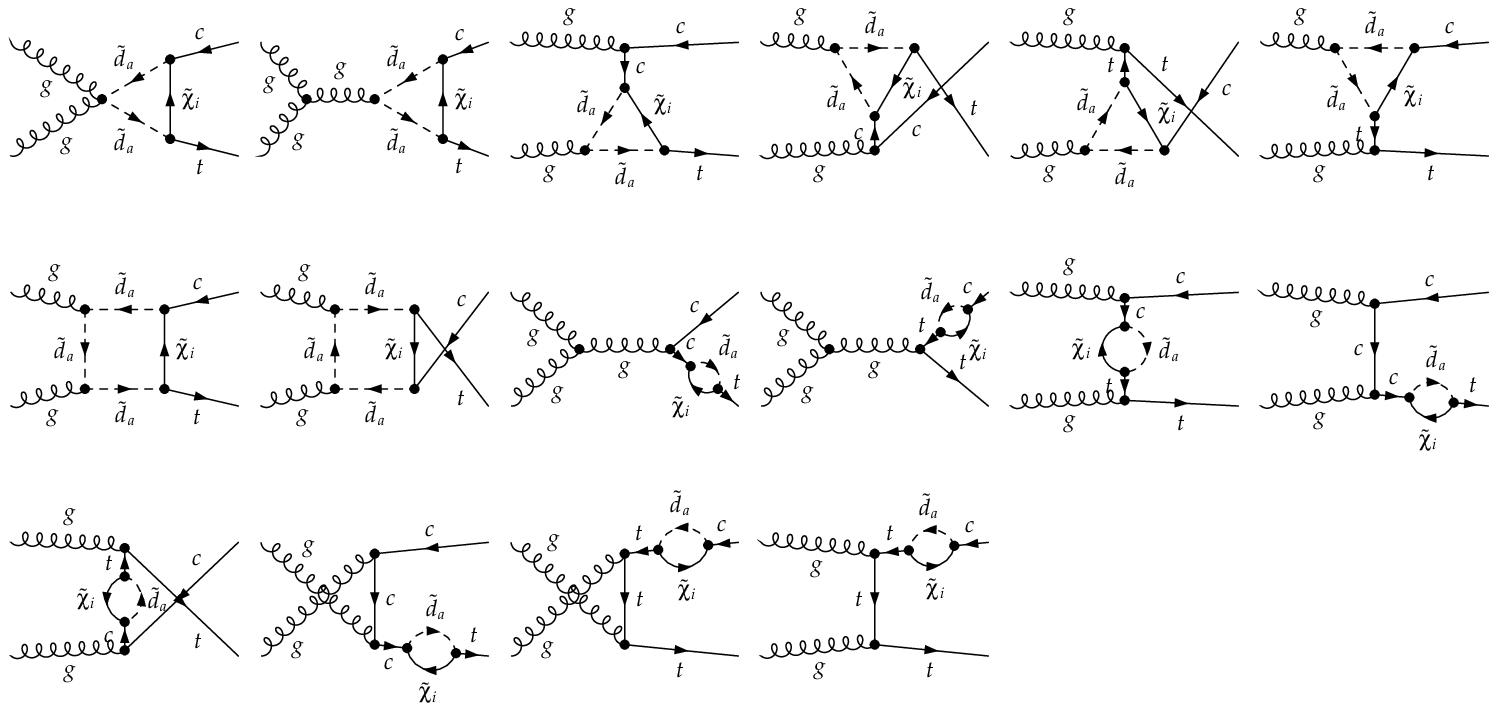}{Feynman diagrams involving chargino and
squark loops for the partonic process $\HepProcess{\Pgluon\Pgluon
\HepTo \Ptop\APcharm}$ in the MSSM. This subset of diagrams is a
part of the charged-current SUSY-EW contribution.\label{fig:dcha}}
\end{center}


\begin{center}
\EPSFIGURE[pt]{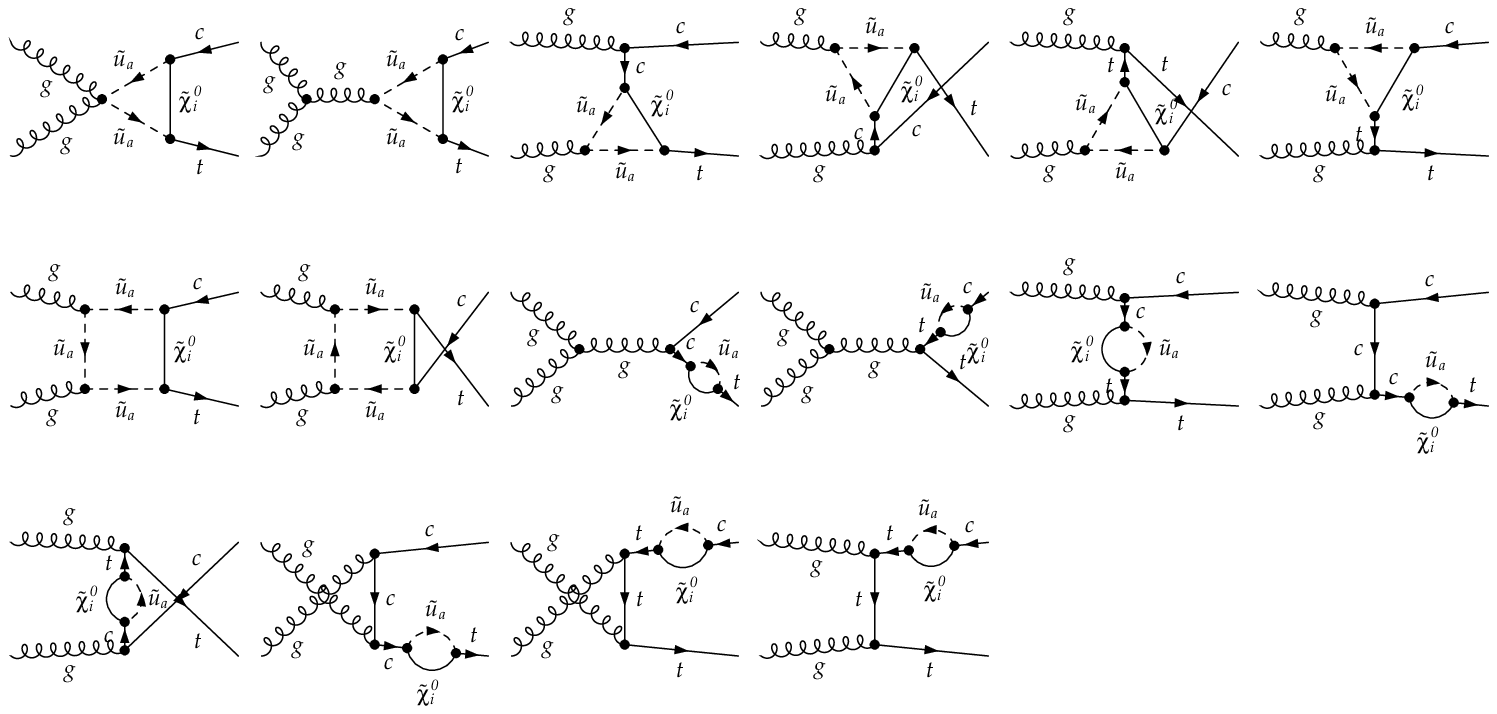}{Set of Feynman diagrams involving
neutralino and squark loops for the partonic process
$\HepProcess{\Pgluon\Pgluon \HepTo \Ptop\APcharm}$ in the MSSM. It
defines the neutral-current SUSY-EW contribution.\label{fig:dneut}}
\end{center}

\begin{center}
\EPSFIGURE[pt]{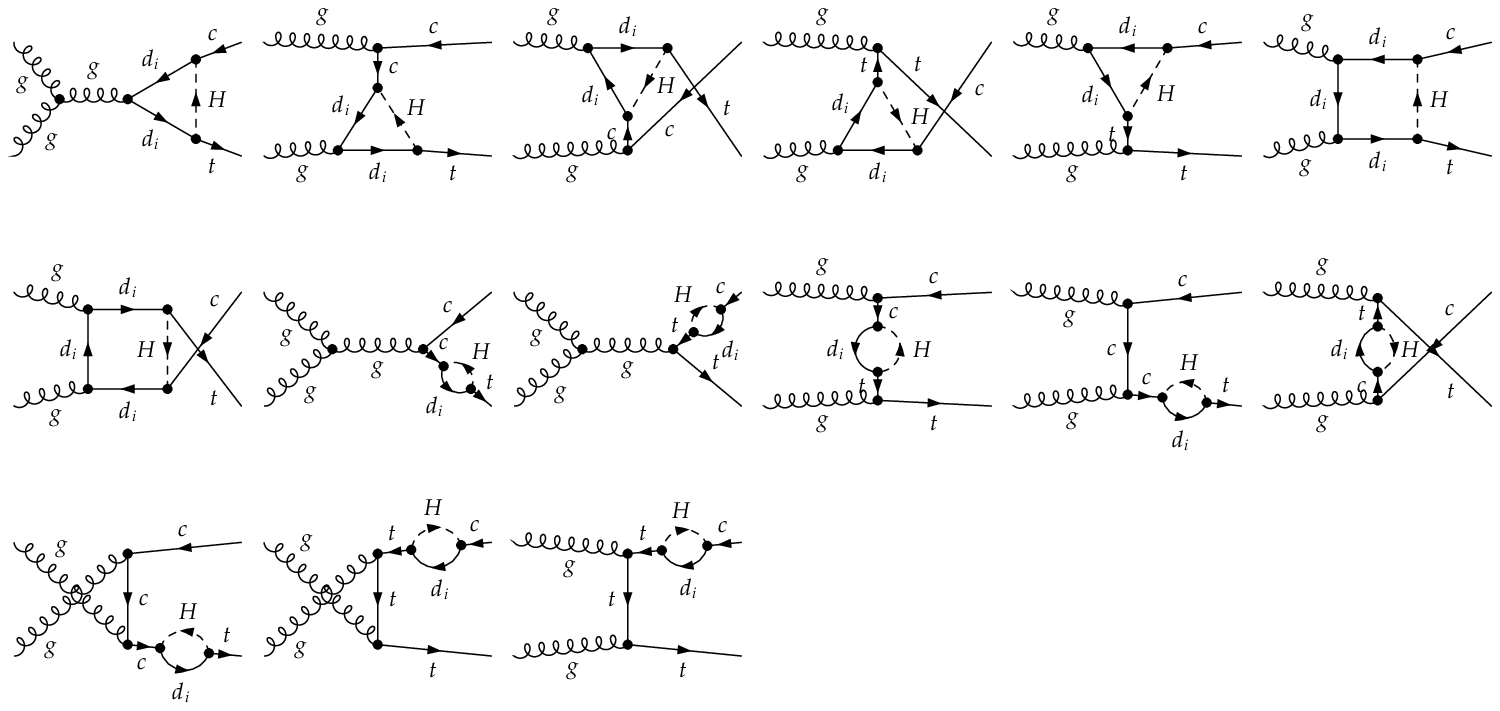}{Feynman diagrams involving charged Higgs
loops for the partonic process $\HepProcess{\Pgluon\Pgluon \HepTo
\Ptop\APcharm}$. Together with the chargino diagrams in
Fig.\,\ref{fig:dcha}, they represent the full non-SM charged-current
contribution to single top quark production in the MSSM.
\label{fig:dhig}}
\end{center}

Regarding the choice of SM parameters, we have taken the heavy-quark
masses and coupling constants given by their corresponding
renormalization group running values at the renormalization scale
$\mu_R$ of the process, see Section~\ref{sec:top}. The running
masses and coupling constants have been explicitly computed by means
of the $\beta$ and $\gamma$ functions at the one-loop level (in the
case of $\alpha_s$, we have made use of an specific subroutine from
the CERNLIB). The obtained values are displayed in Table
\ref{tab:sm}. \vspace{1cm}
\TABLE[pt]{
\centerline{\begin{tabular}{|c|c|c|c|c|c|} %
\hline
$m_c(\mu_R)$ (\GeV) & $m_b(\mu_R)$ (\GeV) & $m_t(\mu_R)$ (\GeV) & $\alpha_s (\mu_R)$ & $\alpha_{em}(\mu_R)$ & $\sin^2\theta_W (\mu_R)$ \\
\hline $0.877$  & $3.024$ & $183.365$ & $0.1170$ & $1/128.89$  &
$0.23$ \\ \hline
\end{tabular}}
\vspace{0.5cm} \caption{Values of the SM parameters at the scale
$\mu_R$ (see the text).\label{tab:sm}} } 
\TABLE[pt]{
\centerline{\begin{tabular}{|c||c|} %
\hline $\tan \beta$ & $5$ \\ \hline $A_t (\GeV)$ & $2238$ \\
\hline $A_b (\GeV)$ & $2000$ \\ \hline $m_{\tilde{g}} (\GeV)$ &
$200$ \\ \hline $M_{SUSY} (\GeV)$ & $746$ \\ \hline $\mu (\GeV)$ &
$400$ \\ \hline $\delta_{23}^{LL}(u)$ & $0.7$ \\ \hline
\end{tabular}}
\caption{Set I of MSSM parameters, that optimizes the SUSY-QCD
contribution in the absence of SUSY-EW effects. \label{tab:sqcd}} }

\vspace{2cm}

Next we evaluate the SUSY-QCD contribution to $\sigma(\processgg)$.
The optimized set of values that we have found for the MSSM
parameters is indicated in Table~\ref{tab:sqcd}. Below we give some
details on their determination, which basically follows the method
of \cite{\main}. However, here we have performed the calculation
with a slightly different set of parameters; in particular, the SM
parameters have been improved by using their RG running values. The
SUSY-QCD corrections are driven by all possible $1$-loop diagrams
(vertex corrections, self-energy insertions and box diagrams)
involving gluinos and squarks (cf. Fig.~\ref{fig:dglu}). In the
following we describe the behavior of the SUSY-QCD contribution to
the total hadronic cross section $\sigma_{t\bar{c}}$ as a function
of a given parameter at fixed values of the others, taking
Table~\ref{tab:sqcd} as a reference. The corresponding results are
reported in Figs.~\ref{fig:sqcd1}, \ref{fig:sqcd2}, \ref{fig:sqcd3}.

To begin with, we consider the curve $\sigma_{\Ptop\APcharm}$ as a
function of $\tan \beta$. What we find is that the cross section
grows steadily until reaching a saturation regime at values of $\tan
\beta \sim 15$. The shaded region is ruled out by  the experimental
determination of the branching ratio $\mathcal{B}_{exp} (\bsg)$\,
\cite{Aubert:2002pd}. The excluded region reflects that the MSSM
calculation of $\sigma_{\Ptop\APcharm}$ in it yields a value of
$\mathcal{B}(\bsg)$ out of the experimental band allowed for this
observable
in the range $\mathcal{B}_{exp} (\bsg)=[2-4.5]\times 10^{-4}$ at the
$3\sigma$-level --  see \,\cite{Aubert:2002pd}  for details. It can
be proven that the overall MSSM amplitude for $\bsg$, and the
purely-SM one, both must have the same sign\,\cite{Gambino:2004mv}.
We have included this restriction also in our numerical codes, so
that we automatically enforce the different scannings over the MSSM
parameter space to be consistent with both the experimental band and
the sign criterion.

We should clarify at this point that up to now we are just retaining
the SUSY-QCD (gluino-mediated) contributions for the computation of
the two observables $\mathcal{B}(\bsg)$ and $\sigma(\processgg)$. In
other words, at the moment we neglect the EW effects both from
supersymmetric particles and Higgs bosons in all these processes as
if they were exactly decoupled. At due time we will switch them on
in combination with the SUSY-QCD effects to evaluate the full MSSM
result.


\FIGURE[t]{

\centerline{
\begin{tabular}{cc}
\resizebox{!}{5.5cm}{\includegraphics{glu_tb_v3.eps}} & \quad
\resizebox{!}{5.5cm}{\includegraphics{glu_at_v3.eps}} \\
(a) & (b)
\end{tabular}}
\caption{SUSY-QCD contribution to the total cross section
$\sigma_{t\bar{c}}$ (in $\picobarn$) and the corresponding number of
events per $100$ \invfb of integrated luminosity at the LHC, as a
function of \textbf{a)} $\tan \beta$ and \textbf{b)} $A_t$ for the
parameters quoted in Table~\ref{tab:sqcd}. The shaded region in
\textbf{a)} is excluded by $\mathcal{B}_{exp} (\bsg)$.
\vspace{0.5cm} \label{fig:sqcd1}}
}

It is worth noting the strong dependence of $\sigma_{\Ptop\APcharm}$
on the trilinear coupling $A_t$ (cf. Fig. \ref{fig:sqcd1}b), where
we have included the approximate constraint $|A_t| \le 3M_{SUSY}$ to
avoid color-breaking minima. We see that $\sigma_{\Ptop\APcharm}$
changes around two orders of magnitude along the explored range. The
dependence of the cross section as a function of the SUSY-breaking
scale ($M_{SUSY}$) and the gluino mass ($m_{\tilde{g}}$) is given in
Figs. \ref{fig:sqcd2}a,b. In both cases $\sigma_{\Ptop\APcharm}$
decreases with the mass scale, as expected from the decoupling
theorem, but this feature is more accentuated with the parameter
$m_{\tilde{g}}$. For instance, $\sigma_{\Ptop\APcharm}$ becomes $10$
times smaller when increasing the gluino mass from $m_{\tilde{g}} =
200$ $\GeV$ to $m_{\tilde{g}} = 500$ \GeV. For the given values of
the parameters in Table~\ref{tab:sqcd}, $M_{SUSY}$ cannot be smaller
that the value indicated there, the reason being that we must
respect the lower mass limits on the squark masses. For the latter
we just take the LEP limits. This means that we do not exclude from
our scanning regions of the parameter space where some physical
squark masses can be as light as $90$ $\GeV$, hence we assume
$m_{\tilde{q}}\gtrsim 90\,GeV$ in our analysis\,\footnote{For a
general overview of the different strategies and up-to-date results
concerning the squark mass bounds, see Ref.~\cite{\PDG}}.

There is also a monotonously decreasing trend when scanning over the
higgsino mixing parameter $\mu$ (Fig.\,\ref{fig:sqcd3}a), although
in this case the variation involves less than one order of magnitude
in the allowed range. Concerning phenomenological bounds, values of
$|\mu|\ge 800$\,$\GeV$ are excluded by the observable
$\mathcal{B}_{exp} (\bsg)$, and also because of the bounds on the
lightest squark mass. LEP bounds also exclude $|\mu| \le 200$\,
$\GeV$ since otherwise the chargino mass limit,
$m_{\tilde{\chi}^\pm} \ge 94$ $\GeV$, would be violated.

The most dramatic dependence of the SUSY-QCD contribution to
$\sigma_{t\bar{c}}$ arises from the explicit flavor-mixing terms.
This can be seen at work in Fig.\,\ref{fig:sqcd3}b, where we scan
over $\delta_{23}^{LL}(u)$. In this case the cross section grows
from $0$ (the gluino-quark-squark coupling vanishes in the limit
$\delta_{23}^{LL}(u)  = 0$) to $2\sigma_{\Ptop\APcharm}\sim 1$
$\picobarn$ at the maximum allowed value of the flavor-mixing, viz.
$\delta_{23}^{LL}(u)\simeq 0.7$ (see below). Larger values of
$\delta_{23}^{LL}(u)$ are excluded by the lower experimental limits
on the squark masses (we recall that the flavor-mixing parameters
participate in the diagonalization of the squark mass matrix, see
Eq.~(\ref{eq:mass2})).

\FIGURE[t]{
\centerline{
\begin{tabular}{cc}
\resizebox{!}{5.5cm}{\includegraphics{glu_msusy_v3.eps}} & \quad
\resizebox{!}{5.5cm}{\includegraphics{glu_mglu_v3.eps}} \\
(a) & (b)
\end{tabular}}
\caption{SUSY-QCD contribution to the total cross section
$\sigma_{t\bar{c}}$ (in $\picobarn$) and the corresponding number of
events per $100$ \invfb of integrated luminosity at the LHC, as a
function of \textbf{a)} $M_{SUSY}$ and \textbf{b)} $m_{\tilde{g}}$
for the parameters quoted in Table~\ref{tab:sqcd}.
\label{fig:sqcd2}}
}
It is important to realize that $\delta_{23}^{LL}(u)$ is also
constrained by $\mathcal{B}_{exp} (\bsg)$. This was advanced in
Section~\ref{sec:formalism}. As we shall next argue, the Set I of
MSSM parameters (cf. Table~\ref{tab:sqcd}) does maximize the
SUSY-QCD contribution to $\sigma_{\Ptop\APcharm}$ within the
present phenomenological constraints on the MSSM parameter space.
We will refrain from writing cumbersome analytical expressions for
the exact formulas. However, we can provide the main analytical
ingredients of the calculation in a schematic way as they will be
useful to understand the physical origin of the SUSY enhancements.
Let us illustrate the procedure by calculating the approximate
optimal value of $\delta_{23}^{LL}(u)$. The starting point in this
discussion is the general form of the SUSY-QCD contributions to
the cross section. It will suffice to consider the partonic
cross-section since all the distinctive dynamical features are
already contained in it. From the formulae of
Section\,\ref{sec:formalism}, and educated guess, we find
\begin{equation}\label{eq:sigmatc}
\sigma (gg\to t\bar{c}) \sim\
\frac{|\delta_{23}(u)|^2}{s}\,\left(\frac{\alpha_s^2}{16\,\pi^2}\right)^2\,
\frac{\mt^2 ({A_t}-\mu/\tb)^2}{\msusy^4}\,.
\end{equation}
Let us briefly explain the origin of some terms in this expression.
There is a loop numerical factor as in (\ref{ordmag}). All
interaction vertices are strong and therefore we must have a
$\alpha_s^4$ dependence. A bit more subtle is the factor of
$\mt\,({A_t}-\mu/\tb)$ at the level of the amplitude, which
originates from the chirality flip of the gluino in the loops and
the corresponding chirality LR transition of the top-squark. This
produces a factor of $M_{LR}^t$ (in the amplitude) emerging from the
stop mass matrix -- see Eq.\,(\ref{eq:mass1}) -- which goes square
in the cross-section. There is of course also the (square of the)
important flavor mixing factor $\delta_{23}(u)$ stemming from
(\ref{eq:mass2}). Finally, the factor $\msusy^2$ in the denominator
of the amplitude (appearing as $\msusy^4$ in the cross-section) is
related to the SUSY particles circulating in the loops. As a matter
of fact, this factor may get contributions from squark masses, the
gluino mass or some combination of them.  By comparison of equations
(\ref{eq:sigmatc}) and (\ref{ordmag}) we can easily deduce the
expected order of magnitude of the SUSY enhancement factor in the
total cross-section. For simplicity, let us approximate
$M_{LR}^t\simeq\mt\,A_t$, which satisfies the conditions of Set I of
parameters in Table \ref{tab:sqcd}. Then the ratio between the SM
and SUSY-QCD partonic cross sections yields
\begin{equation}\label{ratioSMSUSY}
\frac{\sigma(gg\to t\bar{c})_{\rm SM}}{\sigma(gg\to t\bar{c})_{\rm
SUSY-QCD}} \sim\frac{|V_{bc}|^2}{\alpha_s^2|\delta_{23}(u)|^2}\
(G_F\,\msusy^2)^2\ \left(\frac{m_b^2}{m_t^2}\right)\,
\left(\frac{m_b^2}{A_t^2}\right)\,.
\end{equation}
Using the inputs on Table \ref{tab:sqcd} we can easily check that
the SUSY-QCD cross-section is of order $10^7-10^8$ times larger than
the SM one. This result will essentially persist after convoluting
with the parton distribution functions. Therefore, the cross-section
will move some $7$ orders of magnitude upwards, namely from the tiny
value (\ref{SMtc}) up to values of order $1$ pb, which is perhaps
sufficiently large to have a chance of being detected.

In a similar way, we can easily determine the general form of the
branching ratio $\mathcal{B}(\HepProcess{\Pbottom \HepTo \Pstrange
\Pphoton})$ in the MSSM. In this case, we will only indicate the
SUSY factors relevant for our considerations. To start with, a
chirality flip in the bottom-squark line is involved. This follows
from the structure of the Wilson operator describing the effective
low-energy interaction \cite{Bobeth:1999ww}.  Thus, a factor
$({A_b}-\mu\tb)^2$ arises in the corresponding branching ratio. The
result is the following:
\begin{equation}\label{bsg}
B(\bsg) \sim |\delta_{23}{(b)}|^2 \, \frac{\mb^2
({A_b}-\mu\tb)^2}{\msusy^4}\,,
\end{equation}
where the parameter $\delta_{23}{(b)}$ emerges from a mass matrix
similar to (\ref{eq:mass2}) but for the bottom-squark sector. In
this respect, let us recall the relationship (\ref{eq:relsu2})
linking the LL blocks of the stop and sbottom mass matrices, the
only ones involved in this calculation.

Let us define $\delta_{33}^{LR} = m_t {(A_t - {\mu}/{\tan
\beta})}/{M^2_{SUSY}}$, which is nothing but the result of rewriting
the off-diagonal $LR$ matrix element of Eq.~(\ref{eq:mass1}) in
terms of the flavor-mixing parameters defined in
Eq.~(\ref{eq:mass2}). With this definition, we can think of the
curves having a constant value of $\sigma_{t\bar{c}}$ as hyperbolae
in the $\delta_{33}^{LR} - \delta_{23}^{LL}$ plane. Next we take the
up-type squark mass matrix in the following simplified form:
\begin{equation}
  \label{eq:massmatrix}
  {\cal M}^2_{\tilde{q}}=M_{\rm SUSY}^2\left( \begin{array}{c|ccc}
      & c_{\rm L} & t_{\rm L} & t_{\rm R} \\\hline
      c_{\rm L} & 1 & \delta_{23}^{\rm LL} & 0 \\
      t_{\rm L} & \delta_{23}^{\rm LL} & 1 & \delta_{33}^{\rm LR} \\
      t_{\rm R} & 0 &\delta_{33}^{\rm LR} &1
    \end{array}
  \right)\,\,.
\end{equation}
Upon diagonalization, we can see that the allowed values of the
squark masses are such that they fullfill the equation
$(\delta_{23}^{LL})^2 + (\delta_{33}^{LR})^2 =  (1 -
{m^2_{\tilde{q}_1}}/{M^2_{SUSY}})^2$, which is nothing but a
circle of radius $R = 1 - {m^2_{\tilde{q}_1}}/{M^2_{SUSY}}$.
Taking advantage of the above geometrical picture, the only thing
we have to do in order to determine the maximum of
$\sigma_{\Ptop\APcharm}$ is to identify the particular point of
the straight line $\delta_{33}^{LR} = \delta_{23}^{LL}$, in the
$\delta_{33}^{LR} - \delta_{23}^{LL}$ plane, for which the
outermost hyperbola $\sigma_{\Ptop\APcharm} = $const. is tangent
to the circle of radius $R$. Thus, we find that the (approximate)
value of $\delta_{23}^{LL}$ that maximizes
$\sigma_{\Ptop\APcharm}$ reads:
\begin{equation}
  \delta_{23}^{LL}=\frac{\sqrt{2}}{1+\left[1+\frac29\,
  \frac{m_{\tilde{q}_1}^2}{{m_t}^2}\right]^{1/2}}\simeq 0.7\,
\label{eq:maximcomb2},
\end{equation}
where  $m_{\tilde{q}_1}$ has been taken $90$ $\GeV$.  For
comparison, $\delta_{23}^{LL}\simeq 0.68$ if
$m_{\tilde{q}_1}=150\,GeV$, so it does not critically depend on
the lower bound on the squark masses. The parameter value
(\ref{eq:maximcomb2}) is precisely the one quoted in
Table~\ref{tab:sqcd}. The corresponding value for the total cross
section is $\sigma (\processgg) \sim 1$ $\picobarn$, which
translates into a production rate of around $10^5$ events per
$100$ \invfb of integrated luminosity. Notice that the Set I of
MSSM parameters is a convenient choice to maximize the
cross-section since it satisfies $A_b -\mu\, \tan \beta = 0$ and,
as a result, the constraints imposed by $\mathcal{B}_{exp} (\bsg)$
are automatically satisfied and impose no additional restrictions
on $\delta_{23}^{LL}(d)$ and, hence, not on $\delta_{23}^{LL}(u)$
either.

\FIGURE[pt]{
\centerline{
\begin{tabular}{cc}
\resizebox{!}{5.5cm}{\includegraphics{glu_mue_v3.eps}} \quad &
\quad \resizebox{!}{5.5cm}{\includegraphics{glu_mix1_v3.eps}} \\
(a) & (b)
\end{tabular}}
\caption{SUSY-QCD contribution to the total cross section
$\sigma_{t\bar{c}}$ (in $\picobarn$) and the corresponding number of
events per $100$ \invfb of integrated luminosity at the LHC, as a
function of \textbf{a)} $\mu$ and \textbf{b)} $\delta_{23}^{LL}(u)$
for the parameters quoted in Table~\ref{tab:sqcd}. The dashed area
in \textbf{b)} is ruled out by the lightest squark mass bound.
\label{fig:sqcd3}}
}

%
Besides the $\bsg$ bounds, it is also of interest to explore the
impact of the recent double-side constraints on the mass splitting
between the $\PB^0_s$ and $\PaB^0_s$ states \cite{\abazov}, which
yield (at the $90\%$ CL):
\begin{equation}
17\, ps^{-1} < \Delta\,M_s\, < 21\, ps^{-1} \;
\label{eq:value1}
\end{equation}
The former result can be rephrased as follows (see \cite{\mishima}
for details):
\begin{equation}
0.55\, < \Big{|}1 + \frac{\mathcal{M}_{SUSY}}{\mathcal{M}_{SM}}\Big{|} < 1.37
\label{eq:value2},
\end{equation}
\FIGURE[pt]{ \centerline{
\resizebox{!}{5.5cm}{\includegraphics{bmix.eps}}} \caption{$\bmix$
constraints in the $\delta^{LL}_{23}(u)\, - \, m_{\tilde{g}}$ plane.
The present computation takes into account the SUSY-QCD
contributions to the $\bmix$ amplitude {within the mass insertion
approximation} (See e.g. \cite{\gab}. \label{fig:map})}}

\TABLE[pt]{
\centerline{
\begin{tabular}{|c|c|c|}
\hline $\delta^{LL}_{23}(u)$ & $m_{\tilde{g}}\,(\GeV)$  &
$\sigma(\picobarn)$
\\ \hline
$0.70$ & $791$ & $0.00722$ \\ \hline
$0.6$ & $731$ & $0.00422 $ \\ \hline
$0.50$ & $656$ & $0.00332$ \\ \hline
$0.40$ & $556$ & $0.00324$ \\ \hline
$0.30$ & $425$ & $0.00388$ \\ \hline
$0.20$ & $236$ & $0.00732 $ \\ \hline
$0.18$ & $200$ & $0.00806$ \\ \hline
\end{tabular}
} \caption{Total cross section $\sigma(\process)$ (in $\picobarn$)
for the lowest allowed values of the gluino mass $m_{\tilde{g}}$ and
corresponding $\delta^{LL}_{23}(u)$ values in
Fig.\,\protect\ref{fig:map} obtained from the $\bmix$ bounds. The
rest of MSSM parameters are fixed as in Table~\ref{tab:sqcd}.
\label{tab:xsbs}}}

\noindent where $\mathcal{M} = \bra{\PaB^0_s}\ham^{\Delta
B=2}\ket{\PB^0_s}$ stands for the transition matrix element that
describes the $\bmix$ mixing, in which supersymmetric radiative
corrections may certainly play a role. A number of studies
\cite{\bal, \ciuuno, \ciudue} have pointed out the possibility that
those diagrams involving gluino exchange could carry the bulk of the
SUSY contribution to $\mathcal{M}$. It has also been shown that the
aforesaid SUSY corrections are very sensitive to combinations of
flavor-mixing parameters of different chiralities, namely
$\delta^{LL}_{23}(d)\,\delta^{RR}_{23}(d)$,
$\delta^{LR}_{23}(d)\,\delta^{RL}_{23}(d)$. Let us recall, however,
that we are only allowing non-vanishing $LL$-flavor-mixing
parameters, $\delta_{23}^{LL}\neq 0$. Therefore, for consistency, we
will study the influence of the $\bmix$ mixing bounds under this
same assumption. The effective $\Delta B=2$ Hamiltonian describing
the $\PB^0_s-\PaB^0_s$ mixing can be expressed in terms of local
operators and the corresponding Wilson coefficients, see e.g.
~\cite{\bal} for a result {obtained within the so-called mass
insertion approximation (in which the sfermion propagators are
expanded in powers of the flavor-mixing parameters up to the linear
order)}. Focusing on the SUSY-QCD (viz. gluino-mediated)
contribution, and setting to zero all the flavor-mixing parameters
but $\delta_{23}^{LL}(u)$ (and thus $\delta_{23}^{LL}(d)$ through
the $SU(2)_L$ relation (\ref{eq:relsu2})), we are left with a single
Wilson coefficient for the MSSM contribution,
\begin{eqnarray}
C_1^{\tilde{g}}(\mu)\!&=&\!-\frac{\alpha_s^2}{216 m_{\tilde{q}}^2}
\left[ 24 x f_6(x) + 66 \tilde{f}_6(x) \right]
(\delta_{23}^{LL}(d))^2 \label{eq:glu1}.
\end{eqnarray}
This coefficient must be computed at the energy scale where the
actual mixing takes place ($\mu=m_b$) by means of the corresponding
RG equation (Cf. Refs.~\cite{\bal, \gab}).  Finally, we may compute
the amplitude for the $\bmix$ mixing process,
\begin{eqnarray}
\mathcal{M}_{\tilde{g}} (\HepProcess{\PB_s^0 \HepTo \PaB_s^0})&=& \bra{\PaB_s^0}
\ham_{\tilde{g}}^{\Delta B=2}
\ket{\PB_s^0} \nonumber \\
&=&
C_1^{\tilde{g}}(m_b)\,\bra{\PaB_s^0}\tilde{\mathcal{O}}_1\ket{\PB_s^0}
\nonumber \\
&=&
-\frac{1}{3}\,C_1^{\tilde{g}}(m_b)\,m_{B_s}\,f_{B_s}^2\,B_1(\mu),
\label{eq:glu2}
\end{eqnarray}

\noindent where we have used
\begin{eqnarray}
\bra{\PaB_s^0}\tilde{\mathcal{O}}_1\ket{\PB_s^0} &=&
-\,\frac{1}{3}\,m_{B_s}\, f^2_{B_s}\,B_1(\mu)\,, \label{eq:glu3}
\end{eqnarray}

\noindent $B_1$ being a ``bag factor'' (expected to be of order
$\sim 1$), whose numerical value is determined by lattice techniques
\cite{\lat}.

A glimpse at equations ~(\ref{eq:glu1}) and (\ref{eq:glu2}) suggests
that the $\bmix$ mixing bounds may impose further restrictions on
$m_{\tilde{g}}$ and $\delta_{23}^{LL}$. However, we also notice that
a ``bag'' dependence enters the analysis of this system. Namely, the
coefficient $B_1$ is a clear signal of strong binding effects that
cannot be accounted for in the perturbative framework. The
appearance of these coefficients is characteristic of calculations
involving bound state systems by strong interactions. One has to
resort to lattice calculations to estimate them. In this sense, this
kind of calculations are imbued of higher theoretical uncertainties.
Since these bag parameters are not present in the calculation of
${\cal B}(\bsg)$, we deem more suited to treat the respective
constraints derived from $\bsg$ and from $\bmix$ at different levels
of applicability, the former being more solid than the latter. This
way of proceeding also helps to more clearly differentiate the
impact of the different kind of constraints on our FCNC production
process.

Plugging the above expression into our codes, we can single out
which regions in the $m_{\tilde{g}}-\delta_{23}^{LL}(u)$ plane are
excluded (or allowed) by the experimental bounds of
Ref.~\cite{\abazov}. The resulting plot is presented in
Fig.~\ref{fig:map}. As could be expected, only heavy gluino masses
can now be compatible with large flavor-mixing values, whereas light
gluino scenarios enforce $\delta_{23}^{LL}(u)$ to stay low. The
maximum cross sections of our FCNC process for different values of
$m_{\tilde{g}}$ and $\delta_{23}^{LL}$ are presented in
Table~\ref{tab:xsbs}. Notice that the optimal regimes are attained
at the edges of such range, that is to say, when $\delta_{23}^{LL}(u)$ 
is maximum and $m_{\tilde{g}}$ is as
light as permitted by the experimental bounds on the $\bmix$ mixing
-- analogously, when $m_{\tilde{g}}$ is minimum and
$\delta_{23}^{LL}(u)$ is as large as allowed. 
In such optimal regimes we are left with maximum cross
sections barely of the order of $\sigma  \sim 10^{-2}$ $\picobarn$ -
implying \, $\sim 1000$ $\Ptop\APcharm$($\APtop\Pcharm$) pairs per
$100 \,\invfb$ of integrated luminosity. Interestingly enough, such
non-negligible rates could perhaps be attained even for very heavy
gluinos ($m_{\tilde{g}}\sim 800\,\GeV$) in the region of the highest
allowed value of $\delta_{23}^{LL}$,  given by
(\ref{eq:maximcomb2}). {However, we should stress that, in this
range of values of the flavor-mixing parameters, the validity of the
mass insertion approximation could fail significantly. This means
that, in order to estimate with some reliability the
allowed/excluded region of Fig.\, \ref{fig:map}, say for
$\delta_{23}^{LL}\geq 0.5$, it would require the inclusion of
additional terms in the expansion of the sfermion propagators, or
simply to go beyond that approximation through an exact calculation
of the flavor-mixing effects at one-loop. This, more refined,
calculation is at the moment not available in the literature and it
certainly goes beyond the scope of our main aim in this work. On the
other hand, as we have seen, there are other sources of uncertainty
that could challenge the real usefulness of performing such
calculation. }

We conclude that the inclusion of the $\bmix$ mixing constraints
could further limit the SUSY-QCD enhancement capabilities by roughly
$2$ orders of magnitude with respect to the optimal regime in
Table~\ref{tab:sqcd}, {but only if one can apply the restrictions
obtained from the mass insertion approximation for the range of
$\delta_{23}^{LL}$ near the optimal value (\ref{eq:maximcomb2}).}
Recall that the optimal regime (where the relevant FCNC cross
section can be of order of $1$pb) was obtained in agreement with the
experimental restrictions imposed by $\bsg$ alone. However, as
warned above, the $\bsg$ and $\bmix$ constraints are of different
nature and, in our opinion, they should not be placed on equal
footing. The main reason for this is that, for the inclusive decay
$\PB \to X_s \Pphoton$, the theoretical prediction for the branching
ratio can safely proceed by using the Operator Product Expansion
(OPE) within the Heavy Quark Effective Theory. Uncertainties related
to the non-perturbative effects in this case are much lower than
when we are forced to follow other strategies, such as the
introduction of bag factors -- see Eq.~(\ref{eq:glu3}). The latter
must be estimated through lattice calculations and suffer in general
from a larger amount of uncertainty. {This intrinsec source of
uncertainty is part of the reason why we do not consider necessary
for the moment to take into account a more detailed calculation of
the impact of the $\bmix$ mixing constraints on our main process
beyond the mass insertion approximation.}

\vspace{0.2cm}
We turn now our attention to the SUSY-EW contributions to the
process $\processgg$, which we will also treat in combination with
the corresponding effects on the decay $\bsg$. As we have already
discussed in Section~\ref{sec:formalism}, the SUSY-EW effects are
contained in the loop diagrams involving charginos
(Fig.~\ref{fig:dcha}), neutralinos (Fig.~\ref{fig:dneut}) and
charged Higgs bosons (Fig.~\ref{fig:dhig}). It is again of interest
to analize the behavior of these corrections to the total cross
section as a function of the different MSSM parameters. Although we
could try a similar discussion pattern as that employed for the
SUSY-QCD effects, a qualitative treatment of the SUSY-EW effects is
far more difficult and cannot be easily summarized in terms of
approximate analytical formulae as we did for the SUSY-QCD effects.
Therefore, hereafter we limit ourselves to the corresponding
numerical analysis.

\TABLE[pt]{
\centerline{\begin{tabular}{|c||c|} %
\hline $\tan \beta$ & $10$ \\ \hline $A_t (\GeV)$ & $-300$ \\
\hline $A_b (\GeV)$ & $-300$ \\
\hline $A_\tau (\GeV)$ & $-300$ \\
 \hline $m_{\tilde{g}} (\GeV)$ &
$2000$ \\ \hline $M_{SUSY} (\GeV)$ & $250$ \\ \hline $\mu (\GeV)$ &
$400$ \\
\hline $M_1 (\GeV)$ & 48 \\
\hline $M_2 (\GeV)$ & 102 \\
\hline $M_{A^0} (\GeV)$ & 150 \\
\hline $\delta_{23}^{LL}(u)$ & $0.7$ \\ \hline
\end{tabular}}
\caption{Set II of MSSM parameters optimizing the SUSY-EW
contribution in the presence of SUSY-QCD effects. \label{tab:ew}} }
\FIGURE[pt]{
\centerline{
\begin{tabular}{cc}
\resizebox{!}{5.5cm}{\includegraphics{ew_tb.eps}} \; &
\resizebox{!}{5.5cm}{\includegraphics{hig_tb.eps}} \\
(a) & (b) \vspace{1cm}
\end{tabular}}

\begin{tabular}{c}
\resizebox{!}{5.5cm}{\includegraphics{ew_at.eps}} \\
(c)
\end{tabular}
\caption{\textbf{a)} Individual contributions from charginos and
neutralinos, as well as the total supersymmetric electroweak effect
(indicated by SUSY-EW), to the total cross section
$\sigma_{\Ptop\APcharm}$ (in $\picobarn$), and the corresponding
number of events per $100$ $\invfb$ of integrated luminosity at the
LHC, as a function of $\tan \beta$, \textbf{b)} Similar as in (a),
but for the charged Higgs effects only; \textbf{c)} Individual and
total SUSY-EW effects as a function of $A_t$. In all cases the fixed
parameters are as in Table~\ref{tab:ew}. The shaded regions are
excluded by $\mathcal{B}_{exp} (\bsg)$ and the dashed regions are
ruled out by the mass bounds on the lightest supersymmetric
particles. \label{fig:ew1}}
}

In what follows we adopt a second choice of MSSM parameters, Set II
(see Table~\ref{tab:ew}). This choice optimizes the SUSY-EW
contribution. The reason should be clear: Set II is characterized by
a large value of the gluino mass, viz. sufficiently large
($m_{\tilde{g}}\gtrsim 2\,\TeV$) that the SUSY-QCD effects are
virtually decoupled. At the same time, the values of the electroweak
parameters $\mu$, $M_1$ and $M_2$ in Set II insure that some of the
charginos and neutralinos can be relatively light (within their
experimental bounds). The corresponding results are presented in
Figs.~\ref{fig:ew1}, \ref{fig:ew2}, \ref{fig:ew3}. Throughout these
pictures, shaded regions are excluded by the constraints imposed by
$\mathcal{B}_{exp}(\bsg)$, while the dashed areas violate any of the
current mass bounds on the SUSY particles. The respective LEP limits
on the lightest neutralino, chargino and squark read:
$m_{\PSneutralinoOne}\ge 46$ $\GeV$, $m_{\chi^{\pm}_1} \ge 94$
$\GeV$ and $m_{\Psquark_1} \ge 90$ \GeV. We must also insure that
the mass bound on the lightest Higgs boson is respected,
$m_{\PHiggslightzero}\ge 90$ $\GeV$~ \cite{\PDG}.

\FIGURE[pt]{ \centerline{
\begin{tabular}{cc}
\resizebox{!}{5.5cm}{\includegraphics{ew_mue.eps}} \;&
\resizebox{!}{5.5cm}{\includegraphics{ew_msusy.eps}} \\
(a) & (b)
\end{tabular}}
\caption{SUSY-EW contribution to the total cross section
$\sigma_{\Ptop\APcharm}$ (in $\picobarn$) and the corresponding
number of events per $100$ $\invfb$ of integrated luminosity at the
LHC, as a function of \textbf{a)} $\mu$ and \textbf{b)} $M_{SUSY}$
for the parameters quoted in Table~\ref{tab:ew}. Owing to the
$\mathcal{B}(\bsg)$ constraint, the allowed range for $M_{SUSY}$ in
(b) is very narrow (non-shaded area). \label{fig:ew2}}}

We begin our study of the SUSY-EW effects by considering a scan over
$\tan \beta$. The contribution from the charginos (dotted line in
Fig.~\ref{fig:ew1}a) shows a mild growing trend. In the case of the
charged Higgs contribution, the evolution with this parameter
(dashed-dotted line in Fig.~\ref{fig:ew1}b) exhibits a bigger
slope, but the actual size of the contribution remains very small;
in fact, not larger than $10^{-5}$ $\picobarn$ within the allowed
range. The neutralino part slowly decreases with $\tan\beta$ since
the neutralino-quark-squark coupling involves the combination
$C_{\PSneutralino\,q\,\tilde{q}} \sim -|A_t| + {\mu}/{\tan\beta}$,
for $A_t < 0$. The leading SUSY-EW contribution clearly comes from
charginos. Neutralinos give sizeable, but smaller, effects and Higgs
bosons give an entirely negligible yield, typically three orders of
magnitude below that of charginos (at least for the sets of
parameters considered in this work). In this sense, what we usually
call the SUSY-EW contribution (defined as the sum of all the
chargino, neutralino and charged Higgs boson effects) can be
considered virtually identical to the chargino plus neutralino
contributions. In what follows we will avoid plotting the individual
charged Higgs contribution.

Consider next the sensitivity to the trilinear coupling $A_t$
(Fig.~\ref{fig:ew1}c). From the interaction Lagrangians
(\ref{eq:neut}) and (\ref{eq:cha}), it follows that the neutralino
piece is the only one that is sensitive to $A_t$.
Fig.~\ref{fig:ew1}c shows that its dependence is qualitatively
similar to that of the gluino (cf. Fig.\,\ref{fig:sqcd1}b), with a
minimum at low $A_t$ and the largest contributions achieved at the
highest possible values of $|A_t|$. Owing to the $\bsg$
restrictions, the relative sign between $A_t$ and the higgsino mass
parameter $\mu$ becomes of crucial importance. The favored range
corresponds to $A_t\,\mu<0$. For the set of parameters collected in
Table~\ref{tab:ew}, it can be checked that the leading SUSY
contribution to $\mathcal{B}(\bsg)$ is driven by the chargino piece.
This piece is strongly dependent on $\mu$. In particular, the
chargino amplitude accounting for $\bsg$ changes its sign when we
reverse $\mu \rightarrow -\mu$. In our analysis we assume, for
definiteness, $\mu>0$ and $A_t < 0$ (motivated in part by some
preference for the sign $\mu>0$ from $g-2$ of the
muon\,\footnote{For the SUSY effects on $g_{\mu}-2$, see e.g. the
excellent review \cite{Stockinger:2006zn} and references therein.
Let us clarify that the MSSM contribution to this observable depends
on the value of some slepton masses which, however, play a marginal
role in our calculation.}) and scan over the mass parameter $\mu$.
In Fig.\,\ref{fig:ew2}a we can see that there is a mild increasing
behavior of the cross-section for both the chargino and neutralino
contributions. We have also checked that the alternative choice of
signs ($\mu<0, A_t > 0$) does not alter dramatically the leading
SUSY-EW effects.

Next we analyze the behavior of $\sigma_{\Ptop\APcharm}$ under the
variation of  $M_{SUSY}$, i.e. the common squark soft-SUSY breaking
mass scale. Both the chargino and neutralino terms exhibit a
monotonously decreasing behavior as functions of $M_{SUSY}$
(Fig.\,\ref{fig:ew2}b). The neutralinos are the most sensitive ones
and decrease around $3$ orders of magnitude until they reach an
almost saturation regime at a value of $M_{SUSY} \sim 700 \,\GeV$.
Such saturation arises from the particular contributions of the
gaugino-quark-squark couplings in this region of the MSSM parameter
space. Although not shown in the figure, the cross-section
eventually vanishes at very large values of $M_{SUSY}$, as required
by the decoupling theorem. Actually, a large portion of the
$M_{SUSY}$ range in Fig. \ref{fig:ew2}b (viz. $M_{SUSY}\gtrsim288\,
GeV$) lies in the shaded region and, hence, is ruled out by the
$\mathcal{B}(\bsg)$ constraint. Once again we see that the inclusion
of this low-energy observable plays an important role to limit the
scope of the allowed parameter space.

\FIGURE[t]{ \centerline{
\begin{tabular}{cc}
\resizebox{!}{5.5cm}{\includegraphics{ew_m1.eps}}\quad &
\quad \resizebox{!}{5.5cm}{\includegraphics{ew_m2.eps}} \\
(a) & (b)
\end{tabular}}
\caption{SUSY-EW contribution to the total cross section
$\sigma_{t\bar{c}}$ (in $\picobarn$) and the corresponding number of
events per $100$ \invfb of integrated luminosity at the LHC, as a
function of \textbf{a)} $M_1$ and \textbf{b)} $M_2$ for the
parameters quoted in Table~\ref{tab:ew}. \label{fig:ew3}} }

Finally, in Fig.\,\ref{fig:ew3}a,b we explore the response of the
SUSY-EW effects to a variation  with the bino ($M_1$) and wino
($M_2$) mass terms in the gaugino-higgsino mass matrices. While the
chargino piece remains constant, because it is unrelated to $M_1$,
both the chargino and neutralino contributions decrease
significantly with $M_2$. For example, in the case of the charginos,
and for the range explored in Fig.\,\ref{fig:ew3}, it involves a
change of more than $3$ orders of magnitude. Let us note that the
reason for accommodating relatively light SUSY particles in the
spectrum (charginos, neutralinos and squarks) is because we have to
balance the important effects from the relatively light charged
Higgs bosons (${M_{H^\pm}}\gtrsim 170\,GeV$, see Table \ref{tab:ew})
in the SUSY computation of $\mathcal{B}(\bsg)$. This explains why
only a limited upper range of values for $M_{SUSY}$ is allowed in
the current scenario.

Our numerical analysis has clearly identified the region of the
parameter space where the SUSY-EW contributions are most favored and
can compete with the SUSY-QCD effects. To wit: it corresponds to
having very heavy gluinos (of a few TeV) together with light
neutralinos, charginos and squarks (in particular, stops) within the
experimentally allowed lower mass limits. This is indeed the
motivation for having introduced a second choice of MSSM parameters
(Set II in Table~\ref{tab:ew}). Still, it may be useful to consider
a third choice, as we comment below.

In the first part of our study we have considered the SUSY-QCD
effects alone. Afterwards, we have switched on the SUSY-EW effects
in a situation where the SUSY-QCD contribution is subdominant (viz.
characterized by very heavy gluinos). Before closing our
investigation, it may be interesting to consider a third scenario,
namely one in which the SUSY-QCD and SUSY-EW effects are both
present but the former are dominant. (Notice that even when both of
them are favored, the SUSY-QCD effects are usually more important.)
Following our standard procedure, we have numerically determined the
corresponding region of the MSSM parameter space under the
$\mathcal{B}(\bsg)$ constraint. The result is collected in the form
of the parameter Set III in Table~\ref{tab:3}.

In Fig.~\ref{fig:compare}a we plot the cross-section
$\sigma_{t\bar{c}}$ for Set III, i.e. the SUSY-QCD favored one, as a
function of $\delta_{23}^{LL}(u)$. Similarly, in
Fig.~\ref{fig:compare}b we show the yield from the SUSY-EW favored
case, represented by Set II of Table~\ref{tab:ew}. In the two sets
we have the concurrence of SUSY-QCD and SUSY-EW contributions, but
while the former is characterized by relatively light gluinos,
charginos and neutralinos, the latter contains light charginos and
neutralinos, together with very heavy gluinos.
\TABLE[pt]{
\centerline{\begin{tabular}{|c||c|} %
\hline $\tan \beta$ & $6$ \\ \hline $A_t (\GeV)$ & $2200$ \\
\hline $A_b (\GeV)$ & $2000$ \\
\hline $A_\tau (\GeV)$ & $2200$ \\
 \hline $m_{\tilde{g}} (\GeV)$ &
$200$ \\ \hline $M_{SUSY} (\GeV)$ & $750$ \\ \hline $\mu (\GeV)$ &
$-200$ \\
\hline $M_1 (\GeV)$ & 1000 \\
\hline $M_2 (\GeV)$ & 1000 \\
\hline $M_{A^0} (\GeV)$ & 150 \\
\hline $\delta_{23}^{LL}(u)$ & $0.7$ \\ \hline
\end{tabular}}
\caption{Set III of MSSM parameters optimizing the SUSY-QCD part in
the presence of the SUSY-EW contribution and preserving
$\mathcal{B}(\bsg)$ within the experimental bounds. \label{tab:3}} }
\FIGURE[t]{ \centerline{
\begin{tabular}{cc}
\resizebox{!}{8.5cm}{\includegraphics{mixing1.eps} \quad \qquad}  &
\resizebox{!}{8.5cm}{\qquad \quad \includegraphics{mixing2.eps}} \\
(a) & (b)
\end{tabular}}
\caption{Combined SUSY-QCD and SUSY-EW contributions to the total
cross section $\sigma_{t\bar{c}} $ ($\picobarn$) as a function of
$\delta_{23}^{LL}(u)$ for the choices of parameters that optimize
\textbf{a}) the SUSY-QCD term (cf. Table~\ref{tab:3}) and
\textbf{b}) the SUSY-EW term (cf. Table~\ref{tab:ew}). Recall that
$\sigma(\processgg)=2\,\sigma_{t\bar{c}}$. \label{fig:compare}} }

In both panels of Fig.~\ref{fig:compare} we display the individual
contributions from charginos (dotted lines), neutralinos
(dashed-dotted lines) and the gluino (full lines); dashed lines
denote the full SUSY-EW contribution. Finally, the double
dotted-dashed line in this figure accounts for the total sum of all
the $1$-loop diagrams. The fact that the full SUSY-EW contribution
is non-vanishing in the limit $\delta_{23}\rightarrow 0$ is due to
the tiny charged Higgs boson piece. In Fig.~\ref{fig:compare}a the
SUSY-QCD effects are overwhelming as compared to the SUSY-EW ones;
the gluino curve is dominant by at least $4$ orders of magnitude
over the chargino effects (the leading SUSY-EW ones). For this
reason the interference amplitudes between these two kind of SUSY
contributions can be neglected in this case. The highest production
cross-section (namely, the maximum of $2\sigma_{t\bar{c}}$) in this
case can reach $\sim 1$ pb, or equivalently, it implies $\sim
10^{5}$ events per $100$ $\invfb$ of integrated luminosity.

The situation changes dramatically when moving from
Fig.~\ref{fig:compare}\,a to Fig.~\ref{fig:compare}\,b. The latter
shows the results corresponding to the Set II of parameters (the
SUSY-EW favoring ones). The most noticeable feature is that, for
intermediate values of the mixing parameter (say for
$\delta_{23}^{LL}(u)\gtrsim 0.2$), the contributions from the
various sources of SUSY-EW contributions become comparable and are,
within this parameter setup, larger than the SUSY-QCD effects. Owing
to the fact that the chargino interactions involve the
charged-current chargino-quark-squark coupling, the dependence that
we observe on $\delta_{23}^{LL}(u)$ is not clean, i.e. it is not a
direct one; rather, it is obtained via Eq.~(\ref{eq:relsu2}), which
relates the parameters $\delta_{23}^{LL}(u)$ and
$\delta_{23}^{LL}(d)$ through CKM rotation. We have worked out such
relations and implemented them in our codes. Since in this second
scenario the SUSY-EW effects can be comparable to those from
SUSY-QCD, it follows that the interference terms between both sets
of SUSY diagrams cannot be neglected any more. This becomes
transparent by considering the curve in Figure.~\ref{fig:compare}\,b
describing the behavior of the overall $1$-loop contribution. Even
for low values of the mixing parameter, the SUSY-EW corrections turn
out to be significant (for instance, for $\delta_{23}^{LL}(u) \sim
0.4$ the full $1$-loop curve is enhanced by a factor of $2$ with
respect to the SUSY-QCD piece). For $\delta_{23}^{LL}(u) \sim 0.6$
and above, the SUSY-EW part exceeds by roughly one order of
magnitude the SUSY-QCD contribution and becomes the dominant effect
on $\sigma_{t\bar{c}}$. Obviously, for heavier gluinos
($m_{\tilde{g}}$ of a few $\TeV$), the SUSY-EW part would be more
important and it could remain as the only SUSY source of $t\bar{c}$
and $\bar{t}c$ events. In this case, the maximum value of
$2\sigma_{t\bar{c}}$ remains still be sizeable as it would amount to
$\sim 1000$ events per $100$ \invfb of integrated luminosity.

Some observations on previous work on this subject are now in order.
The kind of systematic analysis presented here was not done in
previous work on this subject\,\cite{\Liu,Eilam:2006rb}. Only some
particular regions of the MSSM parameter space were singled out in
these references and, moreover, the important restrictions due to
$\mathcal{B}_{exp} (\bsg)$ have not been taken into account. Thus
e.g. in \cite{\Liu} the bulk of the contribution was missed and the
authors tended to emphasize that the main effects stem from the
mixing in the LR sector of the full mass matrix,
$\mathcal{M}^2_{\tilde{q}}$, namely from $\delta_{ij}^{LR} (i \neq
j)$. Similarly, in \cite{Eilam:2006rb} the various chiral mixing
effects are taken into account and, again, emphasis is made on
scenarios where the maximal effects are obtained from LR mixing.
However, it is not obvious to us whether the important $SU(2)_L$
relation (\ref{eq:relsu2}) was enforced in this reference when
dealing with the LL sector. Furthermore, the simultaneous analysis
of the $\mathcal{B}_{exp}(\bsg)$ within the MSSM is also absent in
\cite{Eilam:2006rb}. We have nevertheless tried to compare our
numerical results with those presented in the latter reference by
adapting our codes to accommodate specific features of their
calculation. Basically, it involves a different choice of SM and
MSSM parameters, a different criterion for the renormalization scale
and the inclusion of some phase space cuts. When doing so, we are
able to successfully reproduce the results provided in Table II of
\cite{\eilamsis}. We should also mention that in \cite{Cao:2007dk}
an analysis is made of FCNC decay modes and the single top quark
production process in which the $\bsg$ constraint is considered.
However, our main emphasis as well as the conditions under which our
results have been obtained, are different. To put it in a nutshell,
in contradistinction to all previous approaches, what we have shown
here goes well along the lines of Ref.\,\cite{\main,JS:RC05} (and
fully generalizes their main conclusion for the complete set of MSSM
quantum effects), to wit: in order to get sizeable SUSY
contributions to $\sigma(\processgg)$ (i.e. cross-sections of order
of $\sim 1\,$pb) compatible with the low-energy FCNC constraints, it
suffices to retain the LL box of the full flavor-mixing matrix,
which is, on the other hand, the only one well motivated by standard
RG arguments. {If, in addition, the $\bmix$ mixing effects are taken
into account in this same kind of scenario, the maximum
cross-section could be further reduced. However, for small values of
$\delta_{23}^{LL}$, in which the mass insertion approximation can be
relied, the reduction is not very significant and applies to a value
of the cross-section which is already small ($\sim 0.01$pb).}

\section{Discussion and Conclusions}
\label{sec:discussion} In this work we have studied the single
top-quark production by strong and electroweak supersymmetric
flavor-changing interactions at the LHC. We have concentrated on the
leading gluon-gluon fusion mechanism, $\processgg$, and restricted
within the context of the Minimal Supersymmetric Standard Model.
Among the supersymmetric flavor-changing interactions in the MSSM,
we have the charged current ones induced by charginos and charged
Higgs bosons, but also the Flavor Changing Neutral Currents (FCNC)
triggered by gluinos and neutralinos. A self-consistent approach to
this high-energy calculation can only be achieved by taking into
account the important experimental constraints on low-energy FCNC
processes, essentially on the radiative $\PB$-meson decays ($\bsg$).
To our knowledge, the first study of the supersymmetric single top
quark production processes at the LHC under these conditions was
performed in \cite{\main,JS:RC05}. Other studies of single top quark
production in the literature\,\cite{\eilamsis,\Liu} simply obviated
the implications on that low-energy process, which is experimentally
well measured and highly sensitive to the dynamics of the SUSY
interactions\,\cite{Bobeth:1999ww,Gambino:2004mv}. However, the
study of \cite{\main} was confined to the computation of the
SUSY-QCD part. Here we have extended it by computing the full
SUSY-EW effects and combined them with the SUSY-QCD ones. In this
way we have obtained the simultaneous MSSM predictions for both
$\sigma(\processgg)$ and $\mathcal{B}(\bsg)$ in full compatibility
with the experimental data on the low-energy process $\bsg$. The
inclusion of the latter is crucial since it carries essential
information on the FCNC parameters in the $\Pbottom$-squark sector.
This information can then be transferred to the LL part of the
$\Ptop$-squark sector via CKM rotation and it thus enters in full
interplay with the study of the production process $\processgg$.

The aforesaid low and high energy interplay is particularly
important in the context of the MSSM, where there are different and
powerful sources of FCNC effects potentially inflecting the final
prediction of
$\mathcal{B}(\bsg)$\,\cite{Bobeth:1999ww,Gambino:2004mv}. For
example, in the electroweak sector, we have the contributions from
charged Higgs bosons and charginos, which have opposite signs and
can have similar impact on the $\bsg$ amplitude. Although the
charged Higgs bosons turn out to contribute very little to
$\processgg$, their effect on $\bsg$ can be very important and,
therefore, must be well balanced against the chargino contributions
which, in contrast, affect the process $\processgg$ in a significant
manner. Only through a careful balance of the SM contribution
against the various (strong and electroweak) supersymmetric effects
(which may carry different signs) is possible to have a spectrum of
relatively light SUSY particles and, at the same time, a safe value
of that branching ratio within the experimental bounds. The ensuing
supersymmetric spectrum obtained in this way, with some of the
particles being relatively light, can then play a fundamental role
to powerfully enhance the single top quark production process
$\processgg$ at the LHC.

Such enhancement has to be measured with respect to the
corresponding standard model (SM) value. To this end, we have
computed the cross-section for $\processgg$ in the SM and found that
it is extremely small ($\sim 10^{-7}$ pb). Since the supersymmetric
effects can increase this result up to $7$ orders of magnitude, it
follows that the process under consideration could have a
cross-section of order of $1$ pb within the MSSM. This situation
would be attained in the most favorable scenario, viz. in the
presence of relatively light gluinos of a few hundred $GeV$ and
large flavor-mixing parameters of order $\delta_{23}\sim 0.5-1.0$.
We note that a cross-section value of $1$ pb amounts to $10^{5}$
events per $100$ $\invfb$ of integrated luminosity.

We should recall that the FCNC Higgs boson decays within the MSSM
($H^0,A^0\to t\bar{c},\bar{t}c$) can also be an additional source of
single top quark events. However, as revealed by the analysis of
\cite{Bejar:2005kv}, these FCNC decay modes are not competitive with
the direct FCNC production process that we have studied here, the
decay mechanism giving rates some two orders of magnitude smaller
under the same set of assumptions. Furthermore, non-supersymmetric
extensions of the Higgs sector of the SM, such as the general
Two-Higgs Doublet Model (2HDM), do not provide any direct production
channel at one-loop. In this context, the only possible source of
$\Ptop\APcharm$ and $\APtop\Pcharm$ events comes from Higgs-boson
decays, which may trigger FCNC decay modes owing to the enhanced
charged Higgs boson interactions. These have been studied in
\cite{\BGST,Arhrib:2004xu} and the result is that this source of
single top quark final states is, again, not competitive with the
direct mechanism $\processgg$ within the MSSM. The latter,
therefore, could be the most efficient source of single top quark
FCNC events in renormalizable perturbartive extensions of the
SM\footnote{Other extensions of the SM, such as topcolor models,
could give
  larger amount of $\Ptop\Pcharm$ pairs at the LHC, see
  e.g. Refs.\cite{Burdman:1999sr,Cao:2002af}.}.

Apart from extending the results of \cite{\main} to include the full
plethora of loop corrections originating from the SUSY-EW sector, in
this work we have improved the calculation of the SUSY-QCD part in
several respects as compared to the previous reference, mainly
through the use of the renormalization group running values of the
SM parameters, and also through an up-to-date set of PDF functions.
Equipped with these updatings, we have compared the SUSY-QCD and
SUSY-EW parts, separately and also in combination, and have
identified and evaluated the region of the parameter space where the
SUSY-EW effects can be sizeable in full compliance with the
restrictions imposed by the low-energy observable
$\mathcal{B}(\bsg)$. That region, which is characterized by
relatively light charginos and neutralinos, could be important,
especially if the SUSY-QCD effects turn out to be suppressed -- e.g.
if the gluinos are very heavy, say, of order of a few TeV. In this
kind of scenarios, obviously controlled by the maximal SUSY-EW
contributions, we find values of $\sigma_{\Ptop\APcharm}$
substantially smaller than those obtained under optimal SUSY-QCD
conditions, but the event rates can still be sizeable (see below).

The following observation is in order. The FCNC effects in the MSSM
may involve, in general, all kinds of chirality mixings (LL, LR and
RR) in the squark mass matrices, and all of them could be important.
This was explicitly shown at the level of FCNC top quark decays in
\cite{Guasch:1999jp}, and confirmed also for top quark production in
\cite{\eilamsis,\Liu}. However, in this work, following and
extending the results of \cite{\main}, we have limited ourselves to
consider flavor mixing in the LL sector only. This is perhaps the
most conservative point of view that one can adopt and, moreover, is
well-motivated by renormalization groups arguments. That restriction
notwithstanding, it is remarkable that the cross-section for
$\processgg$, within the SUSY-QCD favored scenario, can reach the
$\sim$pb  level at the LHC in full compatibility with the
experimental limits on the radiative B-meson decay. This is one of
the most important results that we have obtained.

Another important result is that, if the gluinos would be very heavy
(say, with masses of order of a few TeV) and/or the misalignment of
the quark and squark mass matrices would be negligible (i.e.
$\delta_{23}\simeq 0$), the total MSSM contribution could still be
sizeable thanks to the potential SUSY-EW effects. In these
conditions, the MSSM single top quark production cross-section could
reach $\sim 0.01$ pb. This result, although two orders of magnitude
smaller than the optimal SUSY-QCD one, entails $10^3$ FCNC top quark
events per $100$ \invfb of integrated luminosity at the LHC and,
therefore, it is still sizeable.

{Finally, we have also estimated the possible impact on the FCNC
cross-section under study if we would introduce (on top of the
$\bsg$ constraints) the restrictions induced by the experimental
data on $\bmix$ oscillations. We have found that it could be
significant, namely it could lead to a reduction by roughly two
orders of magnitude of the maximum result, hence a cross-section of
order $10^{-2}$pb at most. However, we emphasize that the
calculation of the SUSY contributions to the $\bmix$ mixing existing
in the literature\,\cite{\gab} has been obtained within the mass
insertion approximation. They are, thus, only valid for sufficiently
small values of $\delta_{23}^{LL}$, namely for values quite away
from our preferred value of this parameter, see
Eq.\,(\ref{eq:maximcomb2}). As a result, our optimized FCNC
cross-section ($\sim 1$pb) need not necessarily undergo the level of
reduction mentioned above. The situation could only be reliably
assessed if we would perform a computation of the SUSY contribution
to $\bmix$ beyond the mass insertion approximation. But this is not
the main aim of our work and, moreover, there are additional,
intrinsic, sources of uncertainty in this calculation which are
associated to our insufficient knowledge of the strongly interacting
matrix element involved in the  $\bmix$ amplitude. These irreducible
uncertainties would remain unsettled. For this reason we have
clearly separated the analysis of the two kind of constraints
imposed on our calculation respectively by the low-energy processes
$\bsg$ and $\bmix$, in the sense that they involve different sort of
assumptions, those from $\bmix$ being susceptible of a higher level
of uncertainty.}

Taking into account the important signature carried by the top quark
and the virtual absence of FCNC contributions both from the SM and
the general 2HDM, the detection and appropriate identification of a
bunch of events of this sort could be interpreted as a hint of SUSY
physics. Let us note that the detection of $\Ptop\Pcharm$-pairs in
the LHC at the level $\lesssim 1 \picobarn$ will be
difficult\cite{Stelzer:1998ni,Han:1998tp}, and that further
experimental studies (including comparison of momentum
distributions, etc.) will be necessary to assess the feasibility of
such measurement. Such studies are beyond the scope of the present
work.

To summarize, SUSY interactions could be an efficient source of
$t\bar{c}$ and $\bar{t}c$ events emerging from direct production
FCNC processes at the LHC. These events, if effectively tagged and
confidently discriminated among the other sources of single top
quark production, could hardly be attributed to any other
alternative FCNC mechanism within renormalizable quantum field
theory. Further studies will be necessary to fully assess this
interesting possibility from the practical point of view, but in the
meanwhile we have shown here that, if we use fairly conservative
assumptions on the theoretical sources of flavor mixing in the MSSM
and consider also the important phenomenological restrictions
imposed by low-energy FCNC processes, there is still a potentially
large supersymmetric enhancement of the FCNC single top quark signal
at the LHC.

\acknowledgments

\noindent The work of DLV has been supported by the MEC FPU grant
Ref. AP2006-00357; JG and JS in part by MEC and FEDER under project
2004-04582-C02-01 and also by DURSI Generalitat de Catalunya under
project 2005SGR00564. The authors are grateful to M. Rauch for
sharing his \textit{HadCalc} program to cross-check our
calculations. We wish to thank W. Hollik and S. Pe\~naranda for
useful discussions in early stages of this work.

\end{document}